\documentclass[%
 reprint,
 amsmath,amssymb,
 aps,
]{revtex4-2}

\usepackage{graphicx}% Include figure files
\usepackage{dcolumn}% Align table columns on decimal point
\usepackage{bm}% bold math
\usepackage{xcolor}
\usepackage{hyperref}
\hypersetup{colorlinks,linkcolor={blue},citecolor={blue},urlcolor={blue}}

\begin{document}

\preprint{APS/123-QED}

%\title{Manuscript Title:\\with Forced Linebreak}% Force line breaks with \\
\title{Formation of Anomalously Energetic Ions
in Hollow Cathode Plume
by Charge Separation Instability}

\author{Yinjian Zhao}
\author{Baisheng Wang}%
\author{Tianhang Meng}
 \email{mength@hit.edu.cn}
\affiliation{
 School of Energy Science and Engineering,
 Harbin Institute of Technology,
 Harbin 150001, People's Republic of China
}

\date{\today}% It is always \today, today,

\begin{abstract}
Hollow cathodes are becoming the bottleneck
of many electric propulsion systems,
because of the sputtering and erosion on both cathodes and thrusters
from the generation of anomalously energetic ions.
So far, it is believed that energetic ions
are formed by waves and instabilities always accompanied in cathode discharge,
but there is no evidence yet that those proposed instabilities can
lead to such high ion energies measured
in experiments.
In this work,
a new mechanism of charge separation instability
in hollow cathode plume
is found via fully kinetic PIC simulations,
which can easily produce energetic ions
to the same level as measured in experiments.
\end{abstract}

\maketitle

%\tableofcontents

A steady working hollow cathode is crucial for coupling with many types of
electric propulsion thrusters,
but hollow cathodes seldom or never operate without oscillations
over a broad band of frequency,
along with the generation of anomalously energetic ions
causing sputtering and erosion
on both cathodes and thrusters.
For example, the cathode life of the early prototype of SPT-100
Hall thrusters had difficulties to reach the 8000 hours lifetime
requirement
\cite{IEPC-95-39};
%\cite{IEPC-93-91,IEPC-95-39};
the PPS-1350 thruster was found to have a more serious asymmetric erosion of
the ceramic channel on the cathode side \cite{10.1063/1.3507308};
in more recent HERMeS thruster tests,
it was also found that the erosion rate is very sensitive to
the cathode position \cite{IEPC-2017-207}.
Therefore, as great progress has been made on thruster performance and stability,
hollow cathodes are becoming the bottleneck
that restricts further improvements of electric propulsion systems.

The phenomenon that hollow cathodes can produce ions
with energies significantly in excess of the discharge
voltage has been found experimentally
since 1990s \cite{doi:10.2514/3.23526}.
Theories have been proposed to explain these
energetic ions, such as the (DC) potential hill model
\cite{kameyama1998potential}
and the ion acoustic turbulence (IAT) model
\cite{10.1063/1.2135409,10.1063/1.2710763}.
%\cite{10.1063/1.2135409,10.1063/1.2710763,doi:10.2514/1.33462}.
Since experimental measurements of the potential
distributions could not find a DC potential hill
responsible for such energetic ions ($>50$ eV) \cite{10.1063/1.2784460},
but the existence of ion acoustic waves has been confirmed in experiments
\cite{PhysRevE.96.023208},
the IAT theory became dominant.
However, as with more and more reports
witnessing the presence of energetic ions under scarce
or undetectable ion acoustic waves,
there seem to be other mechanisms that can contribute to
the existence of such ions.
Especially when it comes to the physical picture of the IAT,
as pointed out by D.M. Goebel et al. \cite{10.1063/5.0051228},
the link between waves and the increase of the ion energy still lacks an explanation,
and there is no evidence yet that IAT alone can lead to such high ion energies.

%However, as claimed by D. M. Goebel et al.
%in 2021 \cite{10.1063/5.0051228},
%the link between waves and the increase
%of the ion energy still lacks an explanation,
%and there is no evidence yet that IAT alone can
%lead to such high ion energies measured
%in experiments that can cause serious sputtering
%and erosion.

From the perspective of numerical simulations
on the cathode instabilities,
fluid models are mainly applied in the channel,
where the plasma density is high and assumed
near equilibrium,
such as zero-dimensional (0D) models
\cite{GURCIULLO2020219}
%\cite{doi:10.2514/6.2019-4246,GURCIULLO2020219}
and 1D models
\cite{panelli2018development}
%\cite{doi:10.2514/2.6146,panelli2018development}
for estimating plasma production and temperature,
and 2D models \cite{10.1063/1.2135409,10.1063/1.2208292}
using the Richardson-Dushman equation,
which can better describe the plasma-wall interaction
and neutral flow effects. There are also hybrid models
\cite{IEPC-2019-865}
representing electrons as fluid and ions as
particles, resulting in favorable comparisons
with some experiments.
In order to simulate the cathode plume,
where the plasma density becomes low,
kinetic approaches are needed.
Pioneers of D. Levko et al. \cite{levko2013two} carried out
2D Cartesian fully kinetic particle-in-cell (PIC)
simulations on a miniaturized orificed hollow cathode,
and Cao et al. \cite{10.1063/1.5029945,Cao_2019}
established a 2D-RZ cylindrical PIC model,
both of which found non-Maxwellian behavior
for electron and ion velocity distributions in the plume.
In addition, K. Hara et al. \cite{Hara_2019}
used a 1D Vlasov solver to study the ion acoustic
turbulence relevant to hollow cathodes.

In order to investigate the instabilities and
energetic ions occurred in the cathode plume,
kinetic models are preferred than fluid models.
While Vlasov solvers are usually limited in
low dimensions due to large computations,
carrying out fully kinetic PIC simulations
without speed-up tricks that may bring in
non-physical effects is also challenging.
Note that D. Levko et al. \cite{levko2013two}
and Cao et al. \cite{10.1063/1.5029945,Cao_2019}
both applied speed-up tricks in their PIC models,
such as reducing
the ion mass, increasing the vacuum permittivity,
or decreasing the simulated cathode size.
These speed-up tricks are known to potentially introduce
%have been shown to
%result in
inaccurate results in Hall thruster
azimuthal instability simulations.
For example, the two 2D classic benchmark works
did not apply any of these speed-up tricks
\cite{Charoy_2019,Villafana_2021},
but attempted to use more
parallel computing techniques to accelerate the simulations.
Because
in the area of Hall thruster simulations,
a great amount of efforts have been devoted to
even carry out massive 3D PIC simulations
\cite{Xie_2024,10.1063/5.0133963},
it is surprising that PIC simulations on hollow cathodes
have not been pushed anywhere close to the limit
of nowadays supercomputing power, taking advantages
of well parallelized PIC codes.

Therefore, in this work, we attempt to
simulate the cathode plume using the fully kinetic PIC
method in 2D-RZ geometry without any speed-up
tricks up to the centimeter scale
and tens of microseconds,
in order to be capable of revealing more
underlying physics with respect to the hallow
cathode instabilities and energetic ions.
The well parallelized PIC code WarpX \cite{warpx}
is applied in this work to carry out
relatively large-scale electrostatic PIC simulations.
Instead of the ion acoustic turbulence,
which would be considered as developed from
a 1$^\text{st}$ order instability,
significant 0$^\text{th}$ order instabilities are observed
in this work,
accompanied with strong oscillating potentials
due to charge non-neutrality.
We therefore would like to call it the
charge separation instability/turbulence,
occurred for example in astrophysical plasmas too \cite{PhysRevD.83.023003}.
The simulation results indicate that energetic ions
can be very easily generated during the evolution of
the charge separation instability.

\begin{figure}
\includegraphics[width=0.48\textwidth]{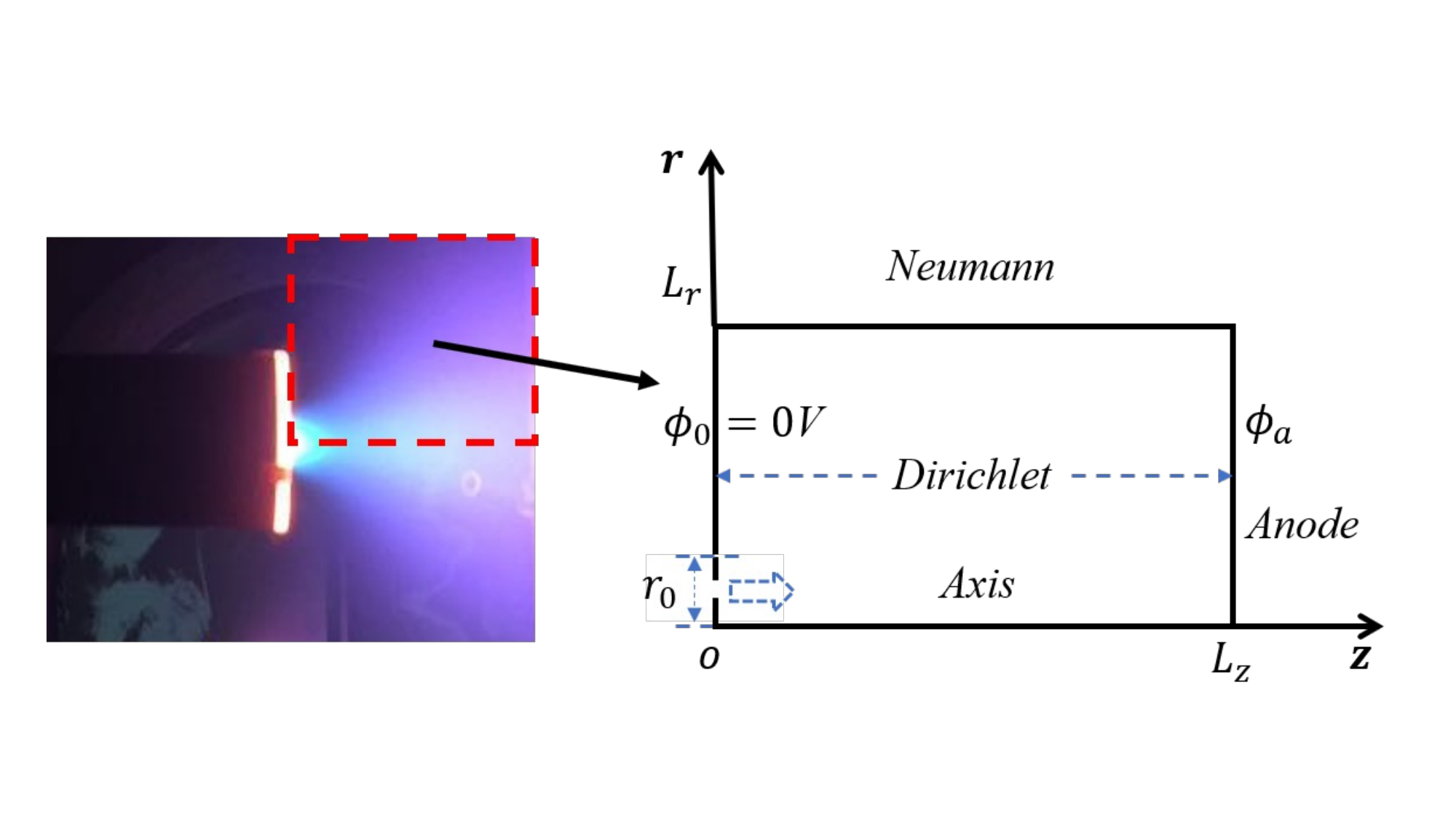}
\caption{A photo of a hollow cathode discharge \cite{Wang_2022} (left)
and 2D-RZ simulation setup (right).\label{fig:setup}}
\end{figure}

The simulation setup is described as follows.
The chosen simulation geometry and domain are illustrated in Fig.\ref{fig:setup},
which is a 2D-RZ cylindrical region with size $1.024\times1.024$ cm,
ranging from the cathode exit to the anode plate.
The Dirichlet condition is applied on
the cathode and anode boundaries
with $\phi_c = 0$ V and $\phi_a = 20$ V, respectively.
The Neumann condition $\partial \phi / \partial r = 0$ is applied on the
maximum $r$ boundary. 
All these three boundaries are set to be absorbing for particles.
At the beginning of the simulation,
electrons and ions start to be injected from the keeper orifice
with radius $r_0=0.5$ mm.
Electrons and ions are sampled according to
Maxwellian velocity distributions
with temperatures $T_e = 5$ eV, $T_i = 0.5$ eV,
and drifting velocities $v_{de}$, $v_{di}$,
respectively.
The corresponding thermal velocities are
$v_{te} \approx 9.4 \times 10^5$ m/s %937768.6310525009
for electrons,
and
$v_{ti} \approx 600$ m/s for the chosen xenon ions. %606.17061564868
Electrons and ions are assumed to have the same number density
at injection $n_0=10^{18}$ m$^{-3}$,
and the electron current is set to be $I_0 = 0.5$ A,
thus the electron drifting velocity can be computed as
$v_{de} = I_0 / (e n_0 \pi r_0^2) \approx 7.9 \times 10^6$ m/s, %7946936.172426809
where $e$ is the elementary charge,
and $v_{di} = 1000$ m/s is chosen empirically.
Based on $n_0$ and $T_e$, the
Debye length is $\lambda_D \approx 1.66 \times 10^{-5}$ m,
such that the cell sizes are chosen to be $\Delta r = \Delta z = 10^{-5}$ m,
resulting in $N_r \times N_z = 1024 \times 1024$ cells.
The time step is set to be
$\Delta t = 0.2 \omega_{pe} \approx 3.55 \times 10^{-12}$ s.
Therefore, to reach a steady state in a few $\mu$s,
more than $10^6$ time steps are needed.
The number of particles injected per time step
(the same for electrons and ions)
is chosen to be $N_p = 400$,
which results in the macro-particle weight
$w_0 = I_0 \Delta t / (e N_p) \approx 27700$,
which is set to be a constant for all macro-particles.
The Monte Carlo collision (MCC) model that accounts for
the electron-neutral elastic, excitation,
and ionization collisions is considered,
with a fixed Xenon atom density distribution,
$n_a = n_{a0}/[(r^2+z^2)^{1/2}+1]^2$,
where the atom density at the cathode exit 
is set to be $n_{a0}=10^{19}$ m$^{-3}$,
and the atom temperature is 0.5 eV.
This simulation is carried out using
256 MPI ranks on
a single computing node with two AMD EPYC 9754 CPUs.
The run time is about 3.39 days
to finish $5\times 10^6$ time steps,
i.e., 17.75 $\mu$s.

\begin{figure*}
\includegraphics[height=2.7cm]{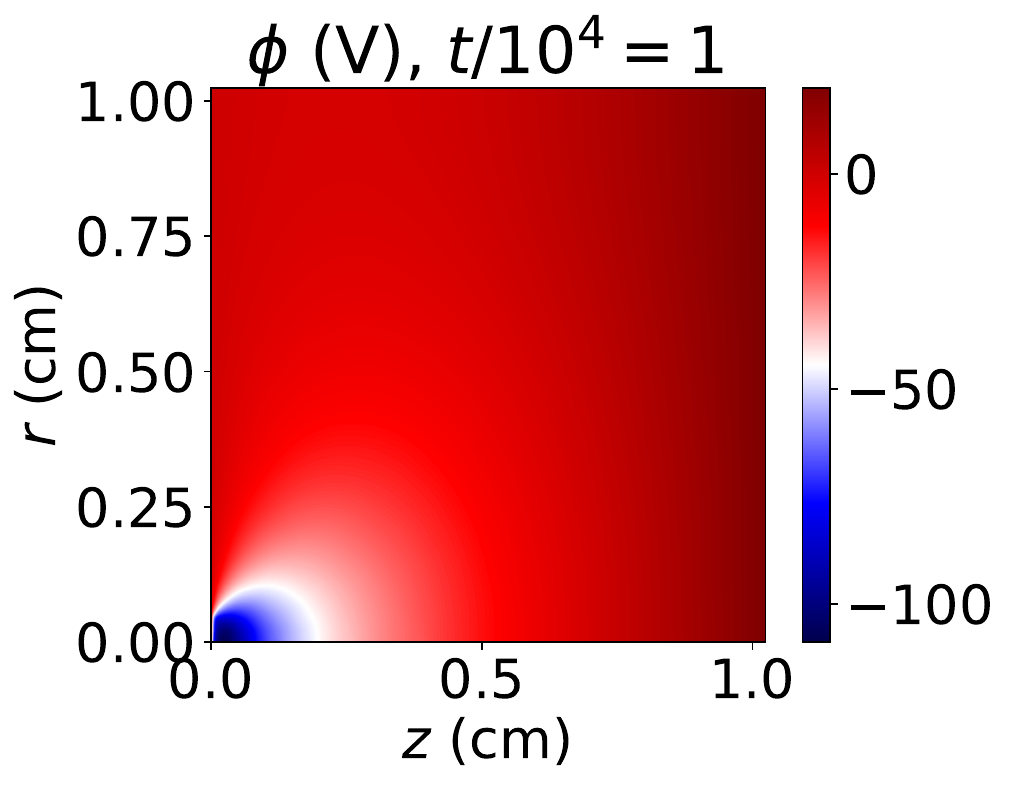}
\includegraphics[height=2.7cm]{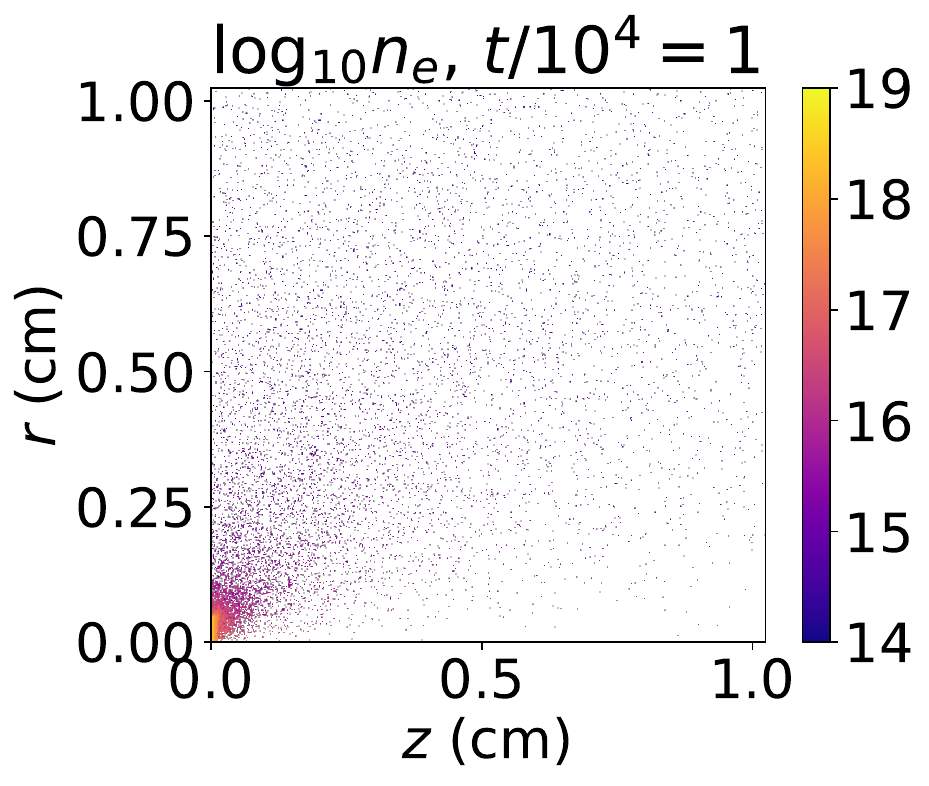}
\includegraphics[height=2.7cm]{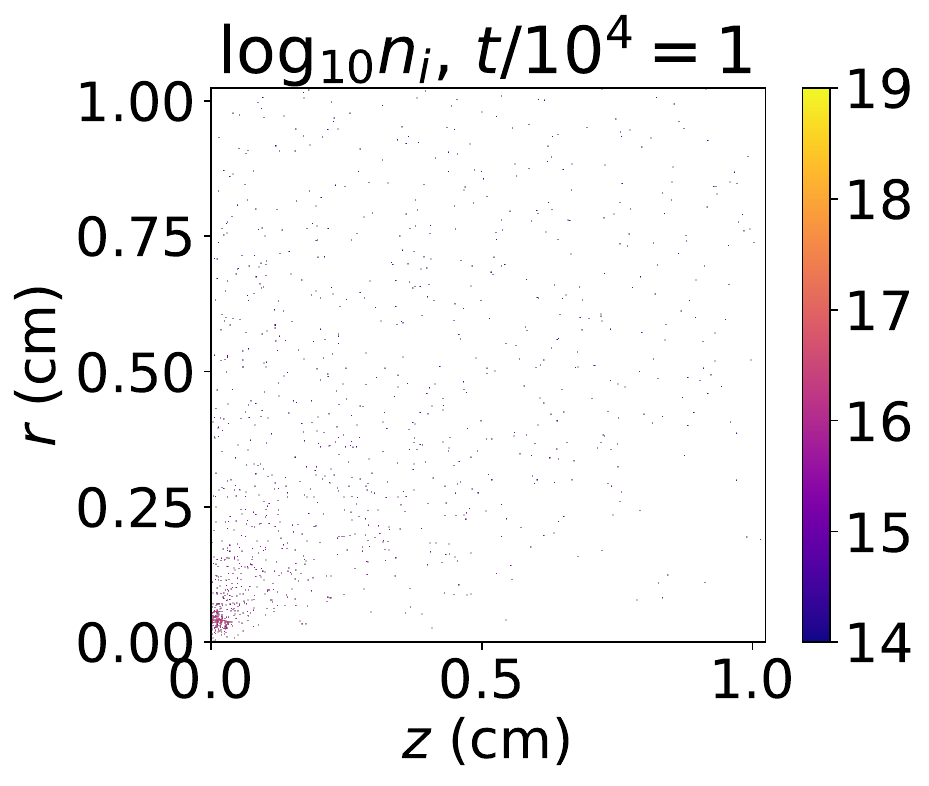}
\includegraphics[height=2.7cm]{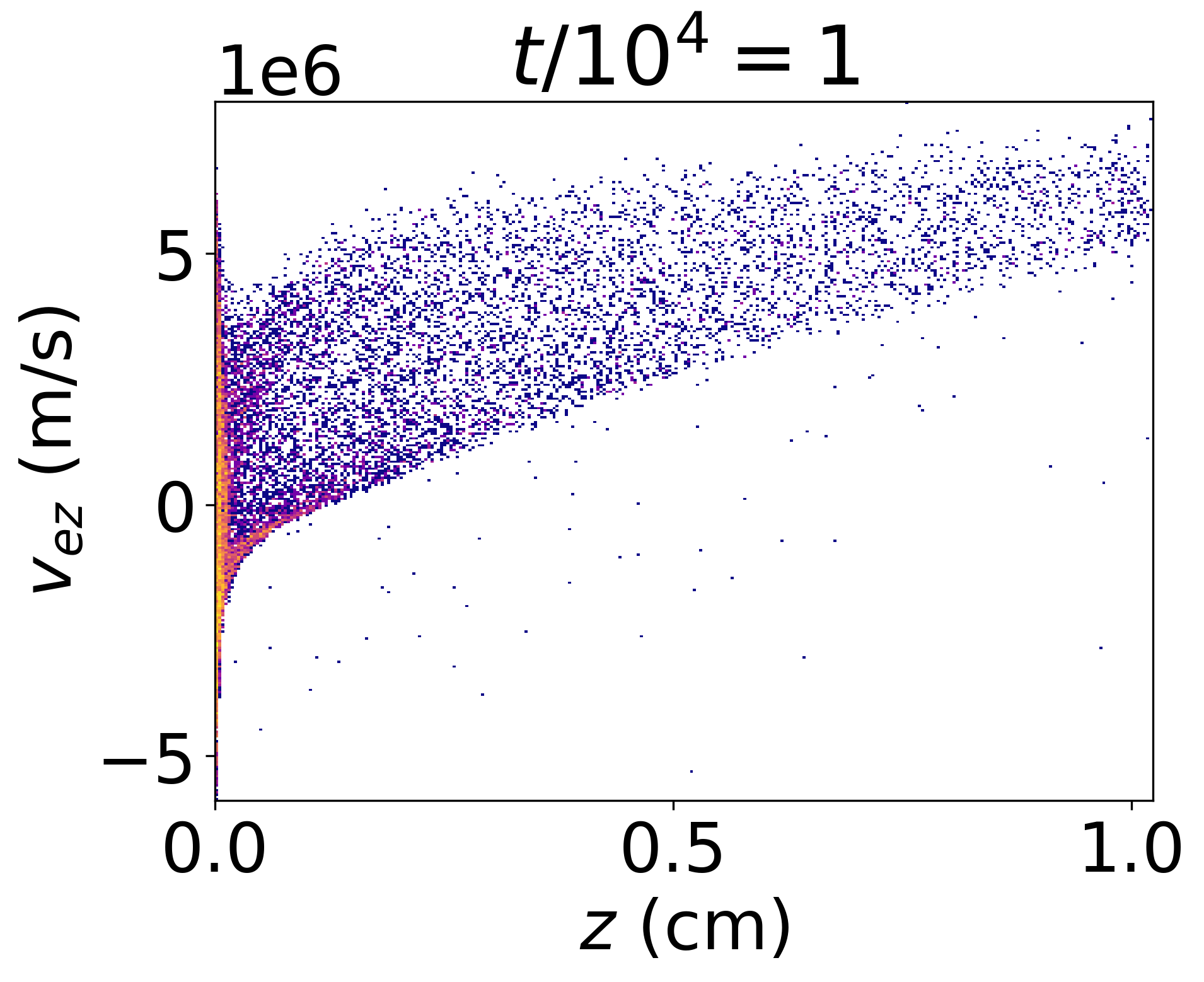}
\includegraphics[height=2.7cm]{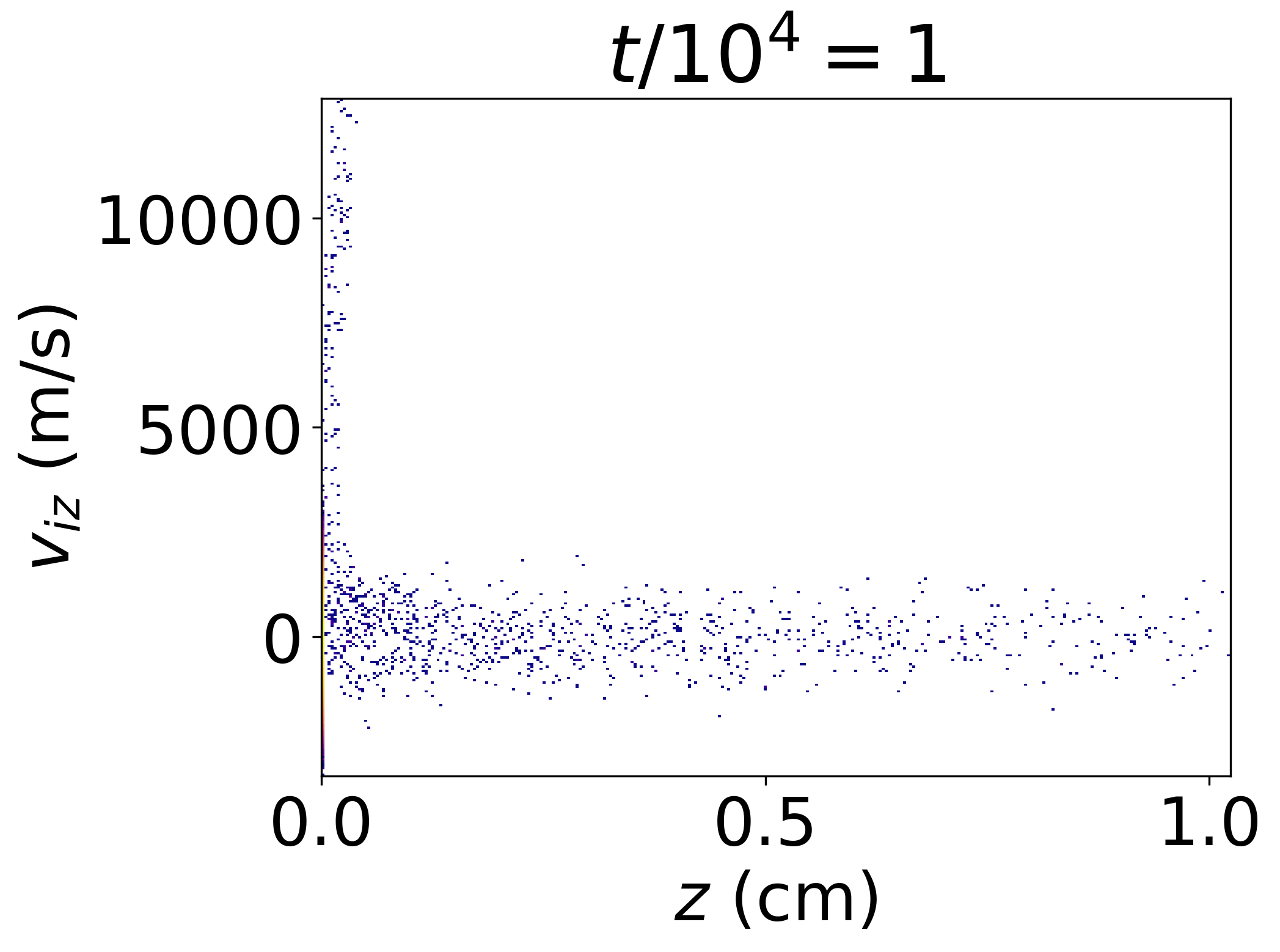}\\
\includegraphics[height=2.7cm]{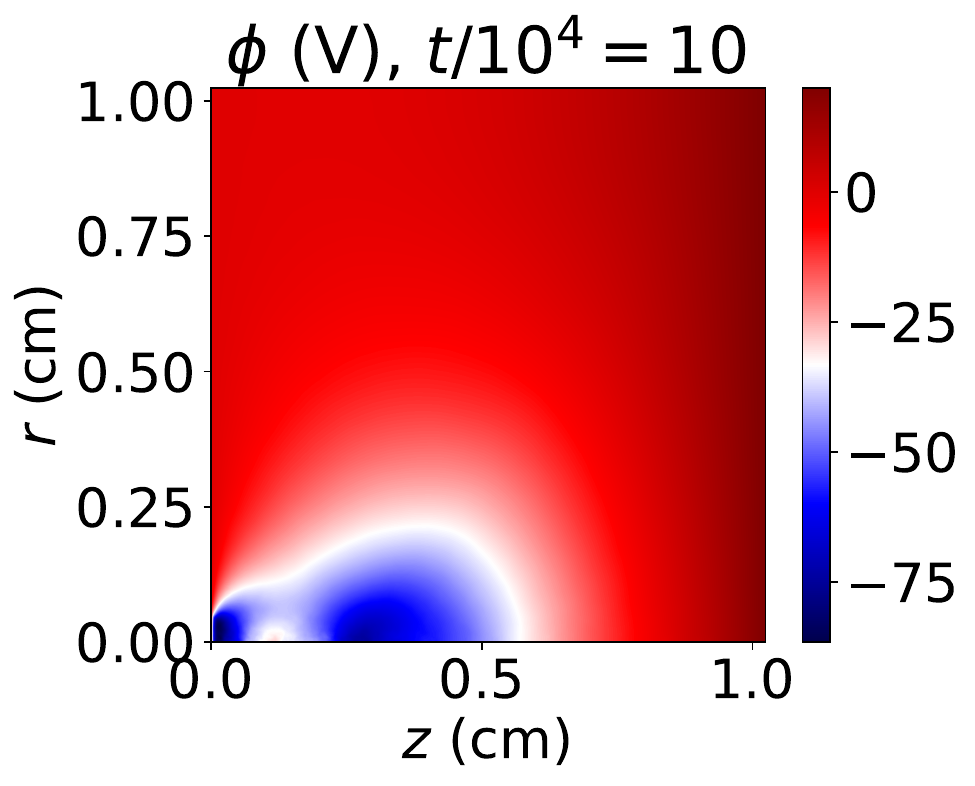}
\includegraphics[height=2.7cm]{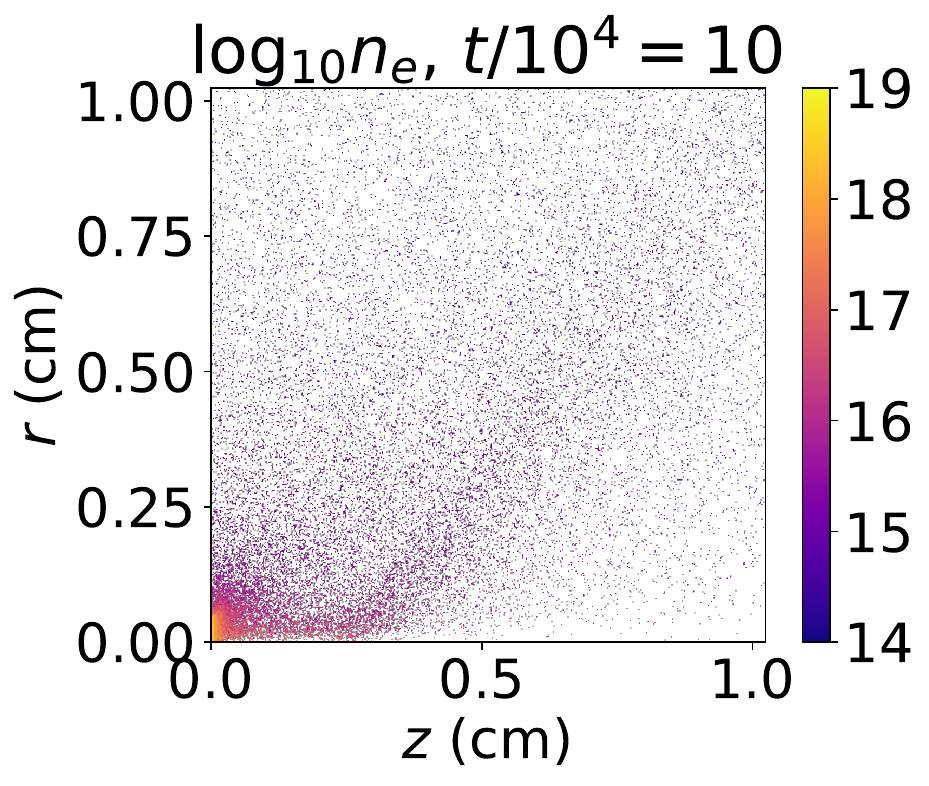}
\includegraphics[height=2.7cm]{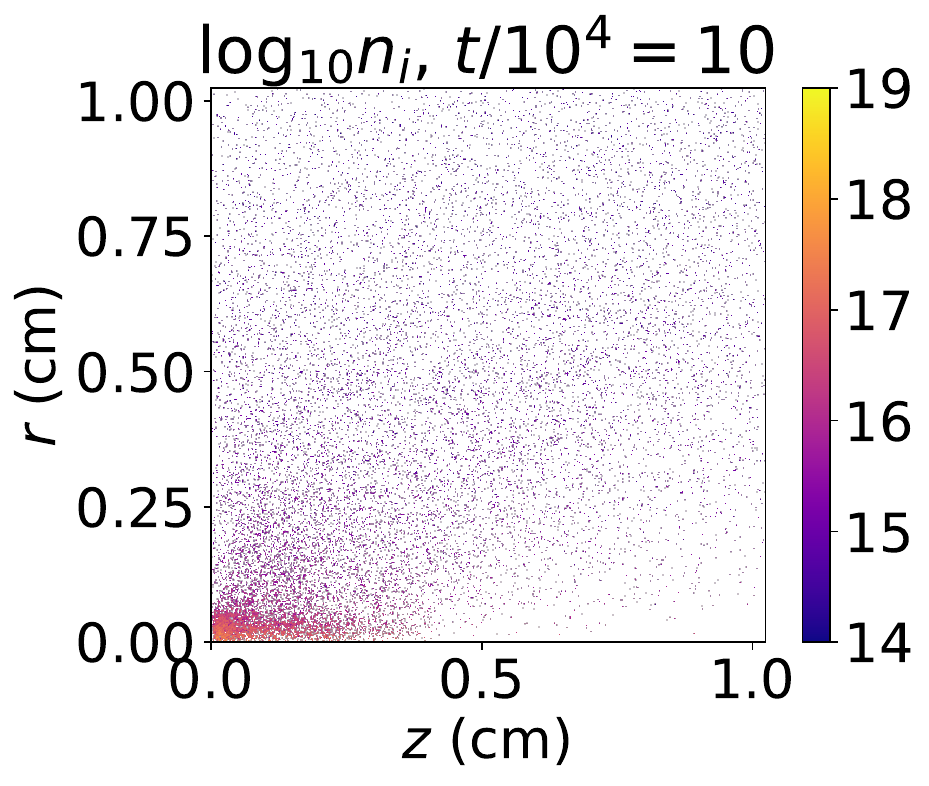}
\includegraphics[height=2.7cm]{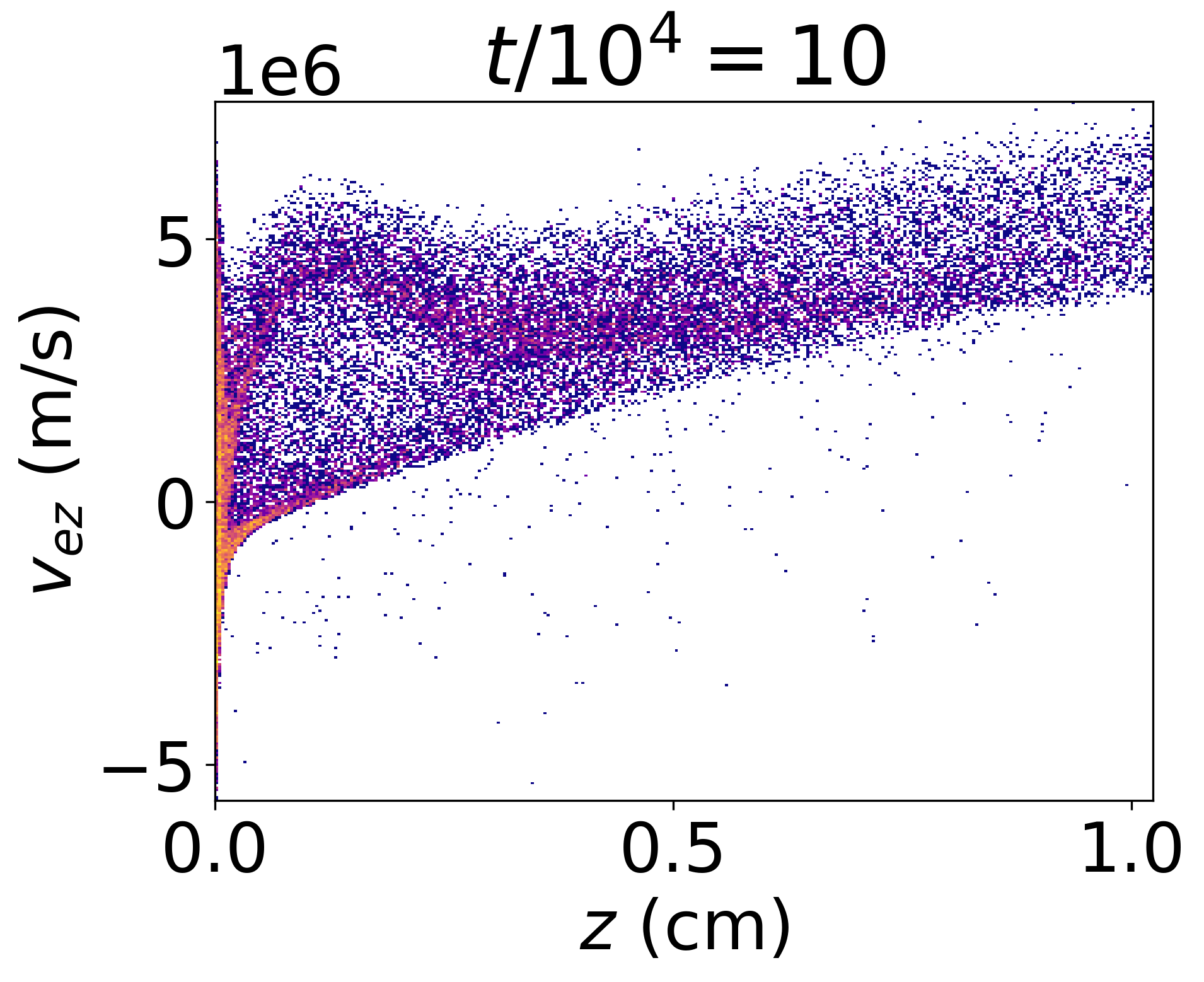}
\includegraphics[height=2.7cm]{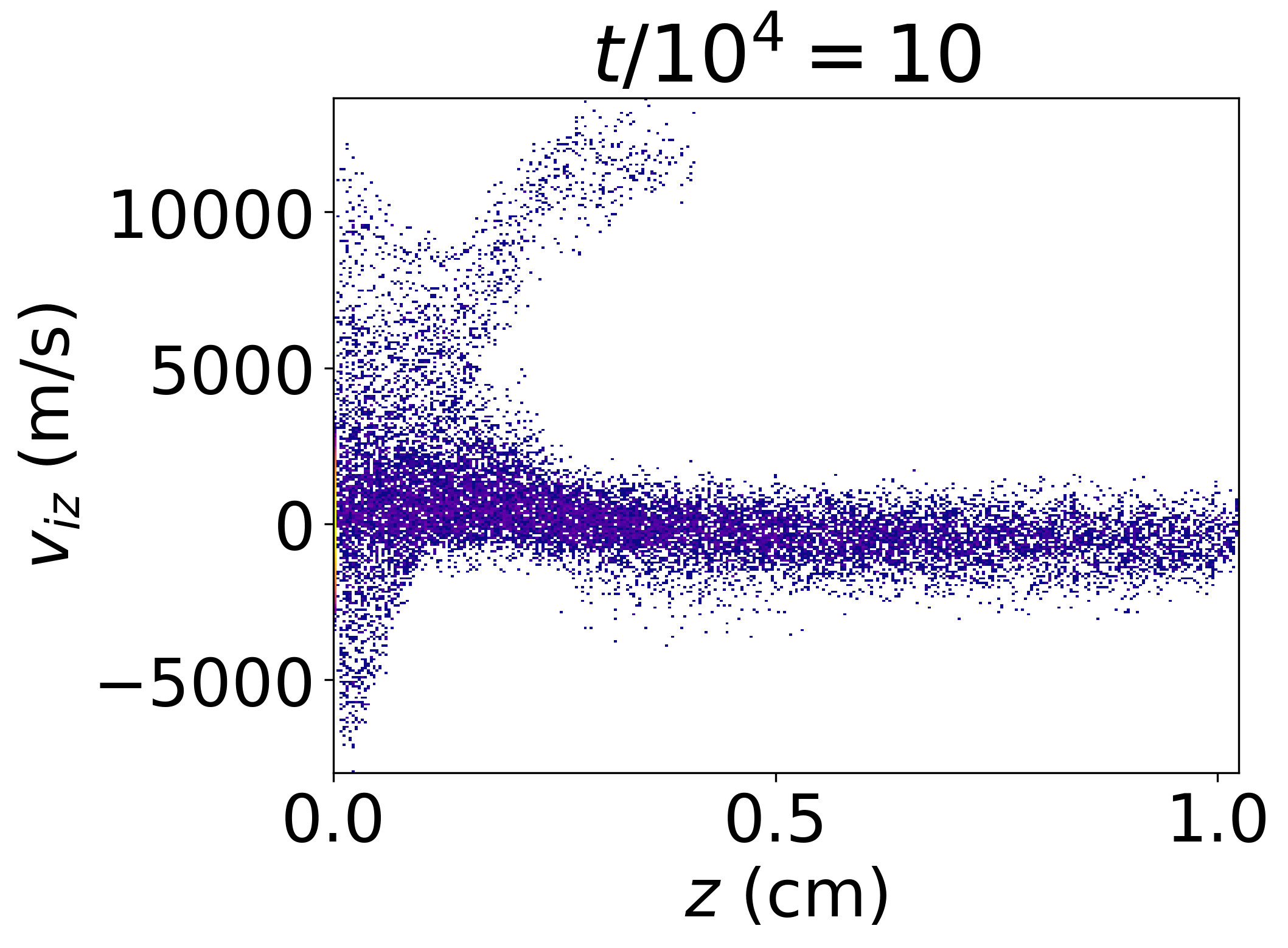}\\
\includegraphics[height=2.7cm]{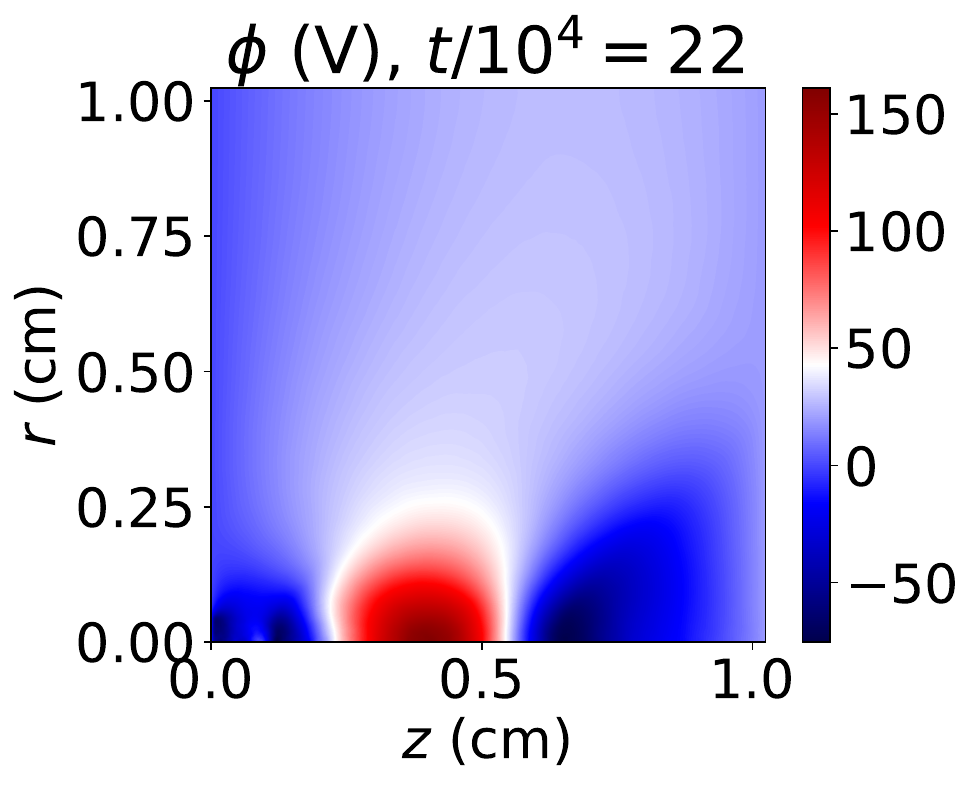}
\includegraphics[height=2.7cm]{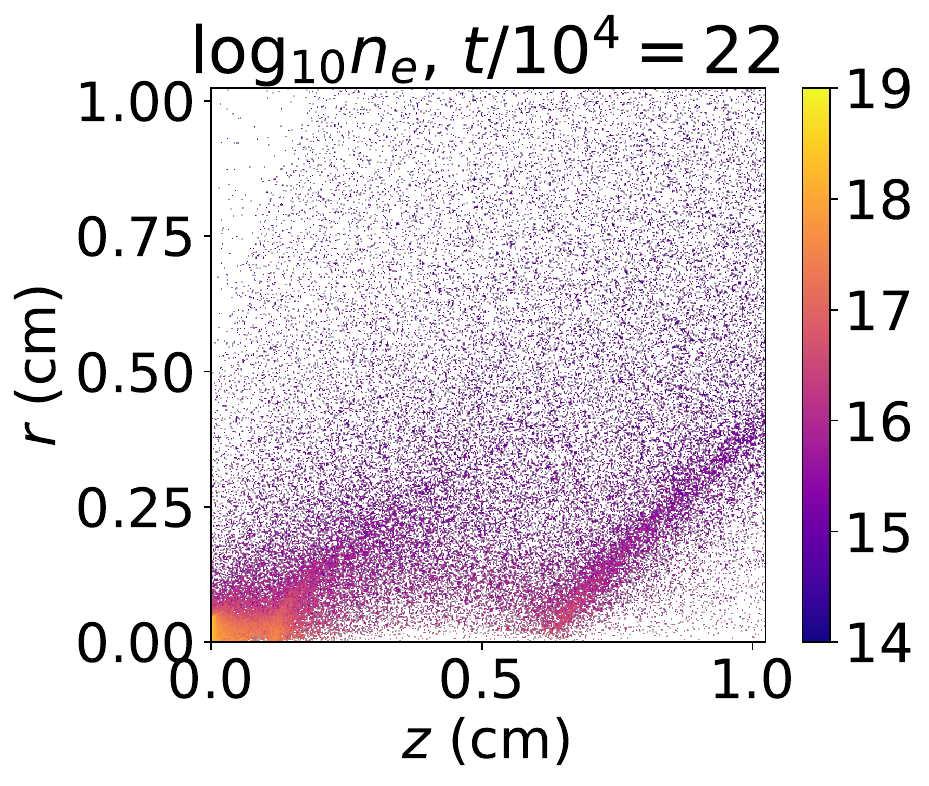}
\includegraphics[height=2.7cm]{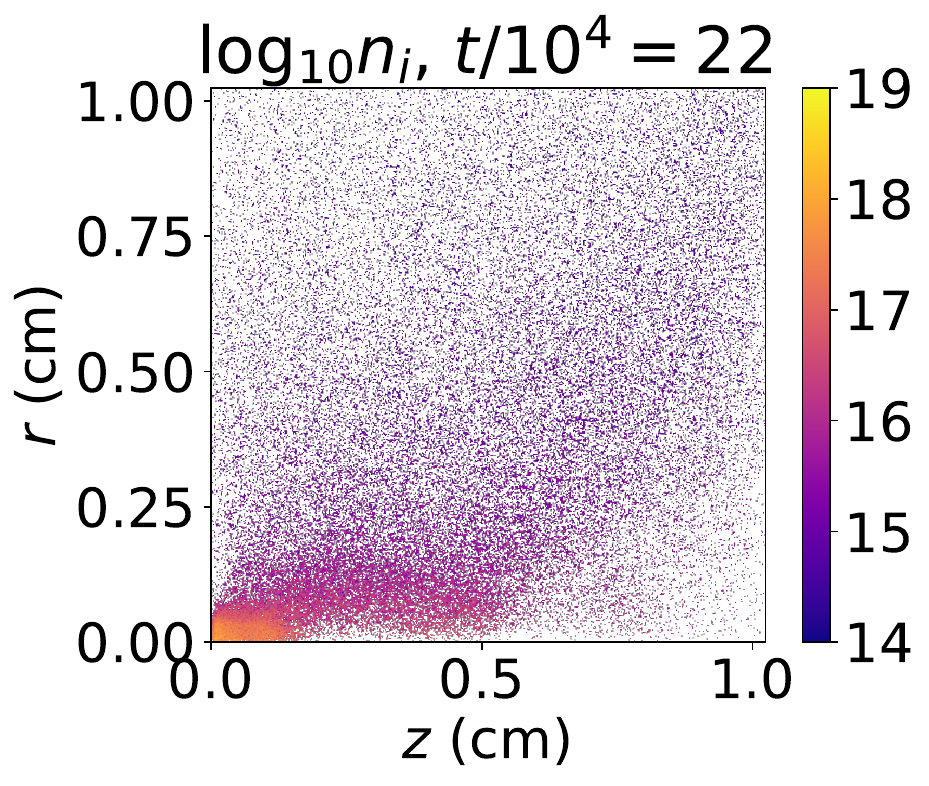}
\includegraphics[height=2.7cm]{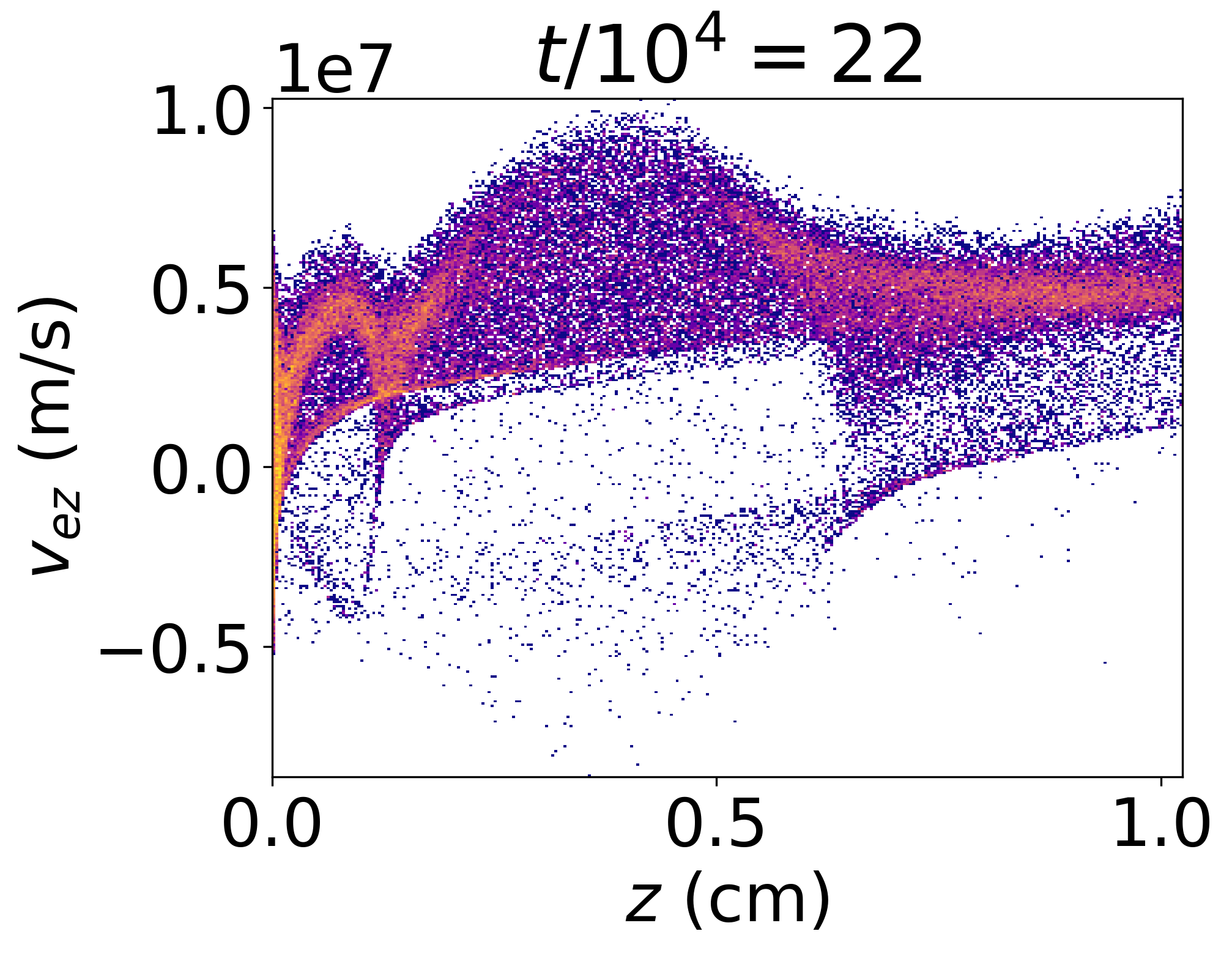}
\includegraphics[height=2.7cm]{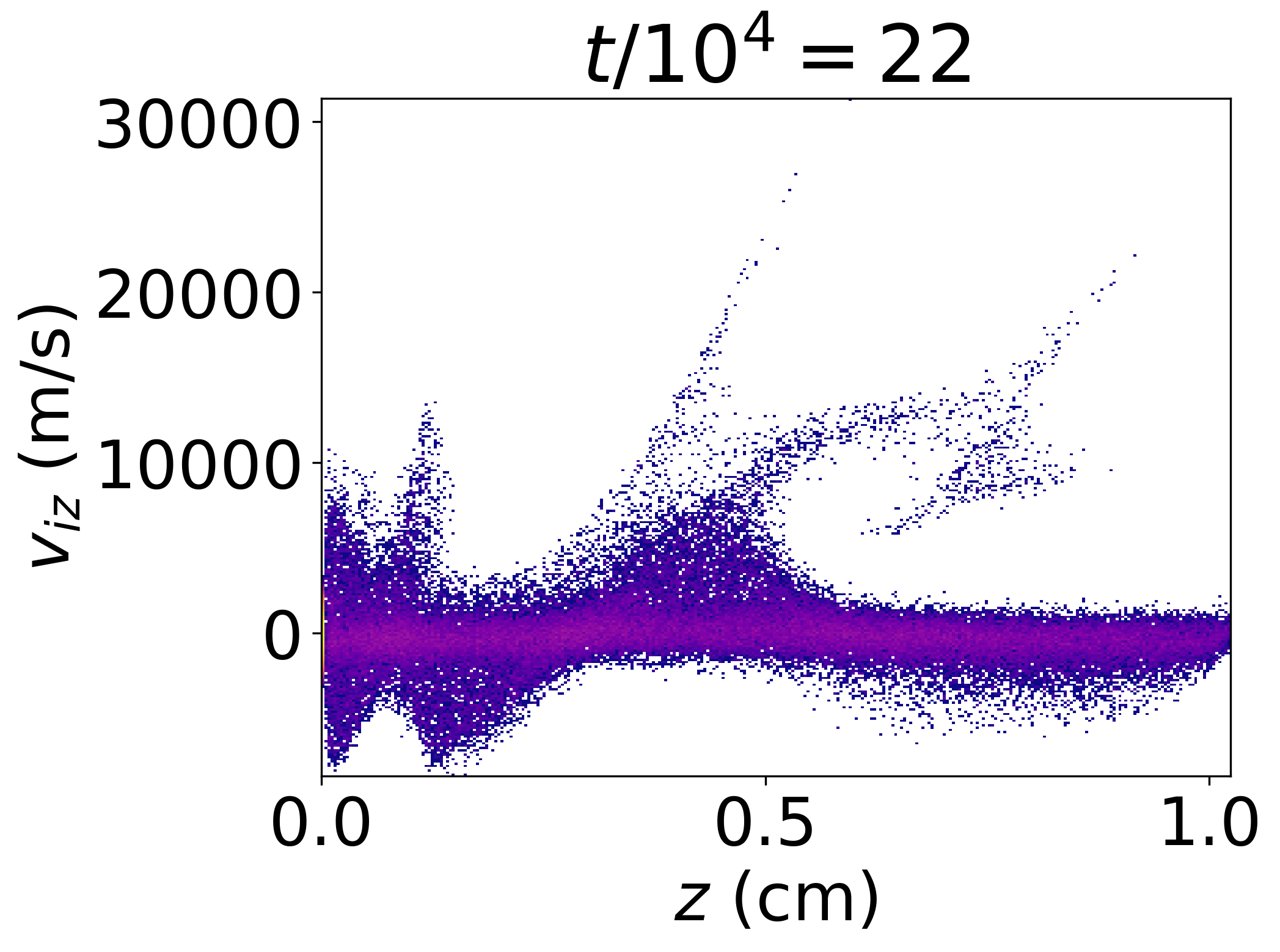}\\
\includegraphics[height=2.7cm]{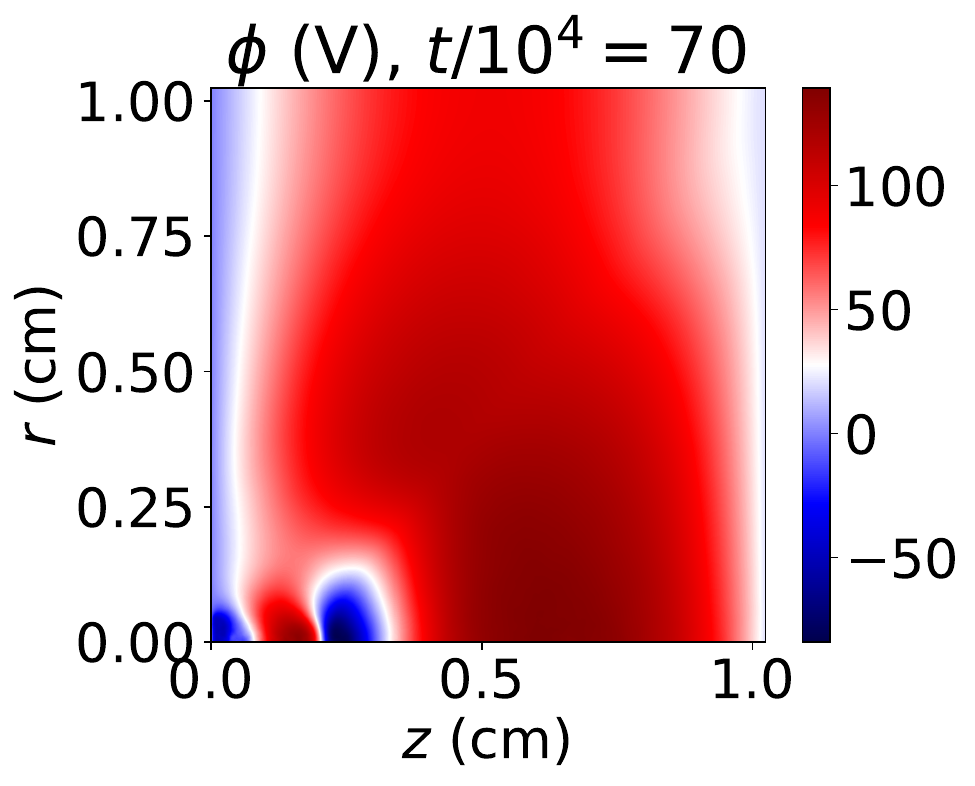}
\includegraphics[height=2.7cm]{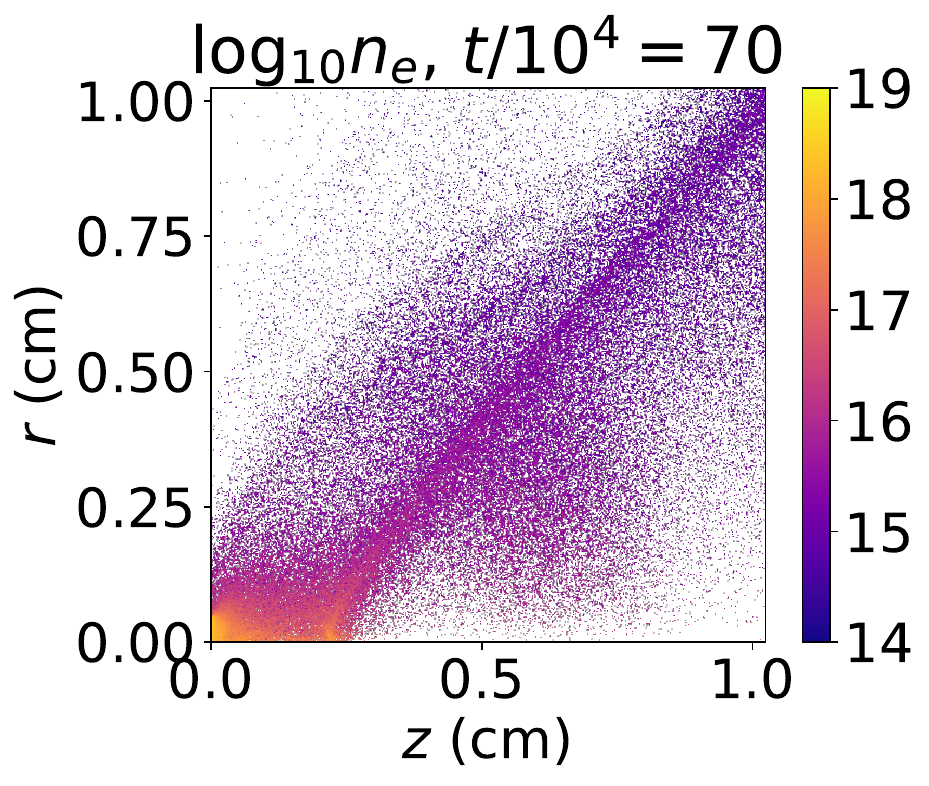}
\includegraphics[height=2.7cm]{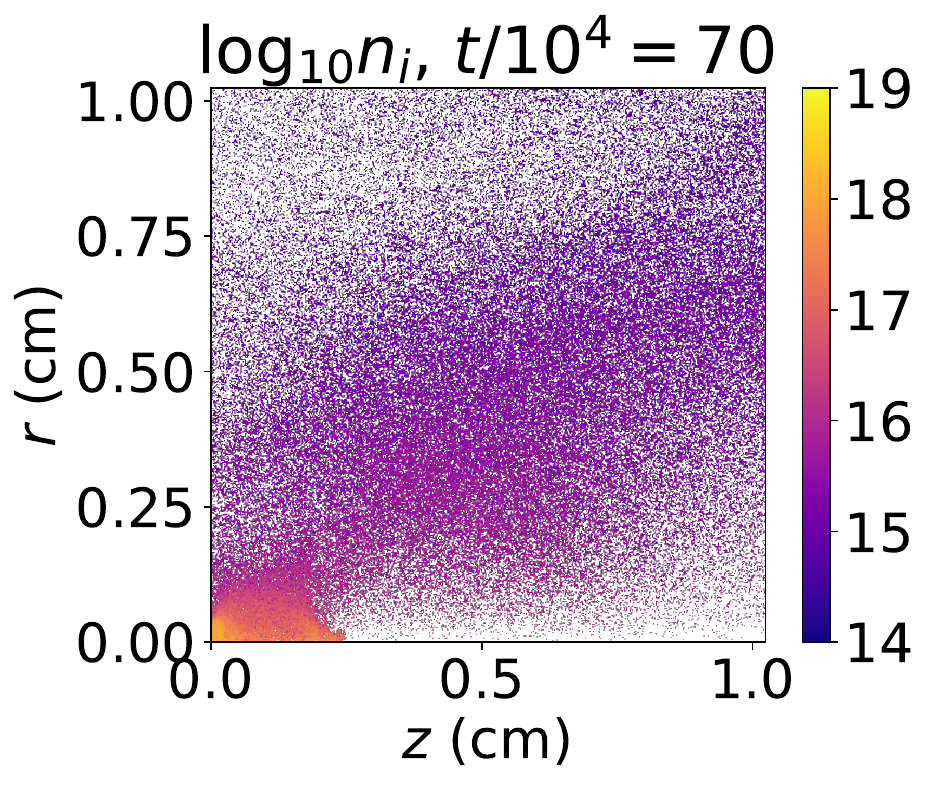}
\includegraphics[height=2.7cm]{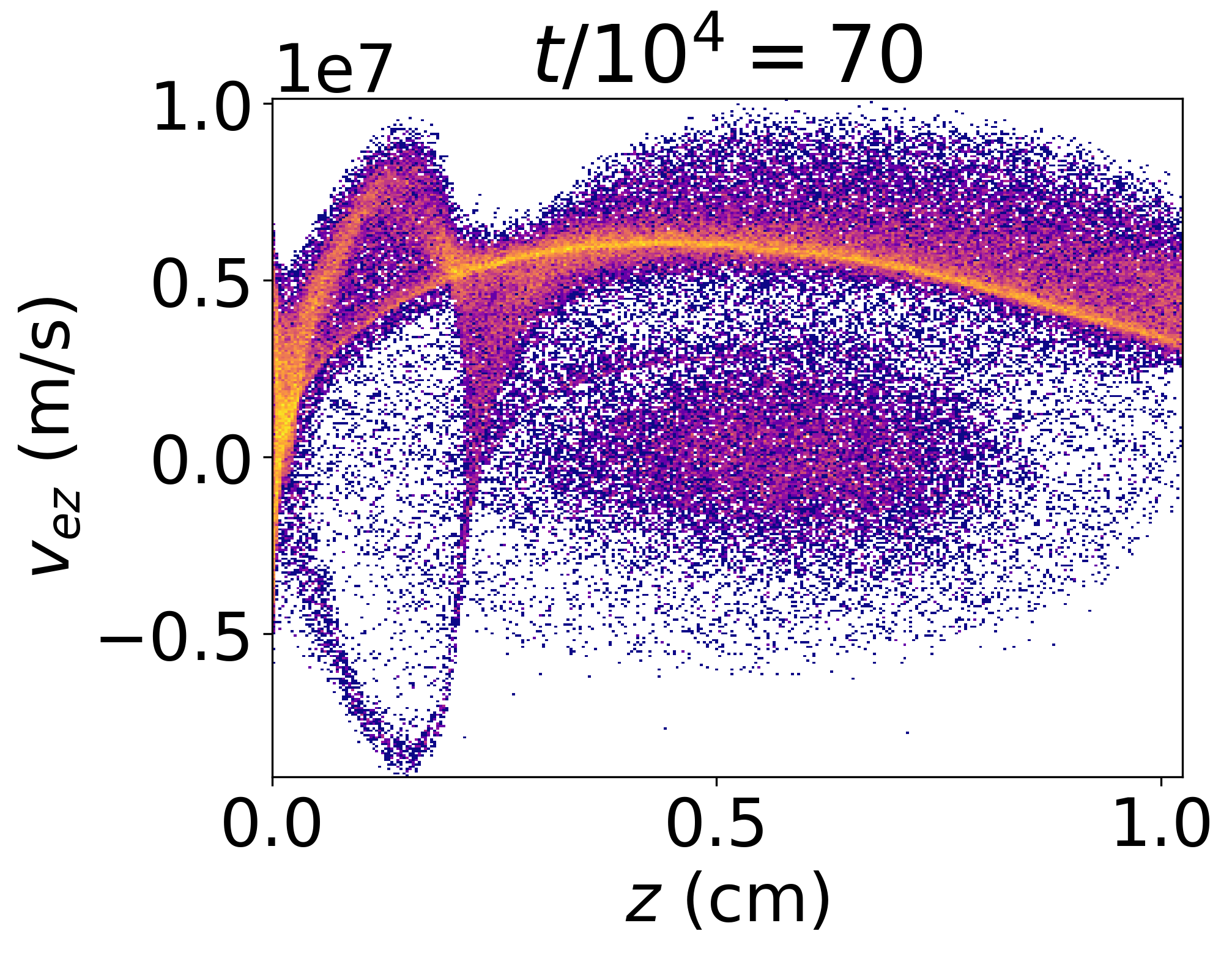}
\includegraphics[height=2.7cm]{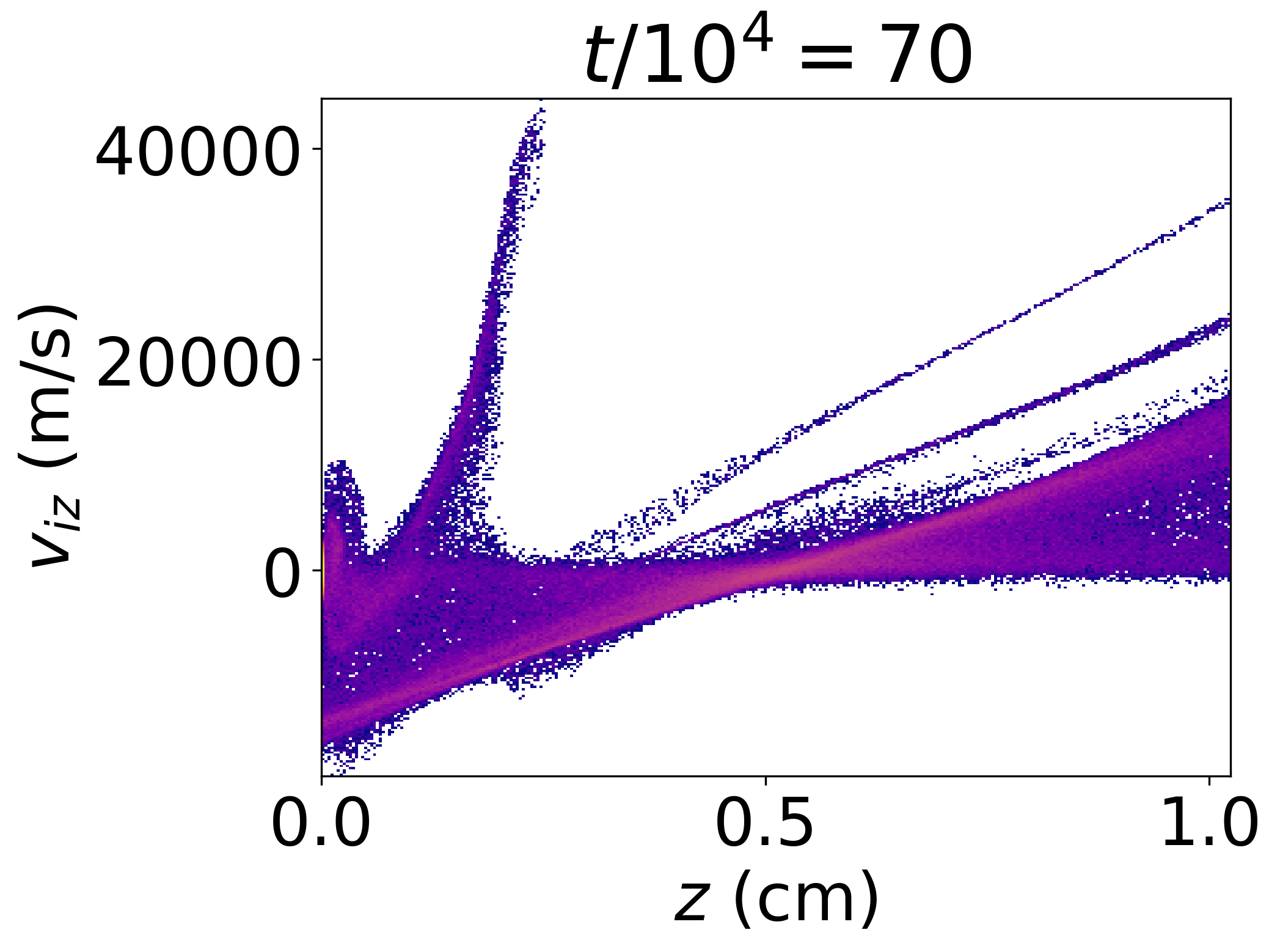}\\
\includegraphics[height=2.7cm]{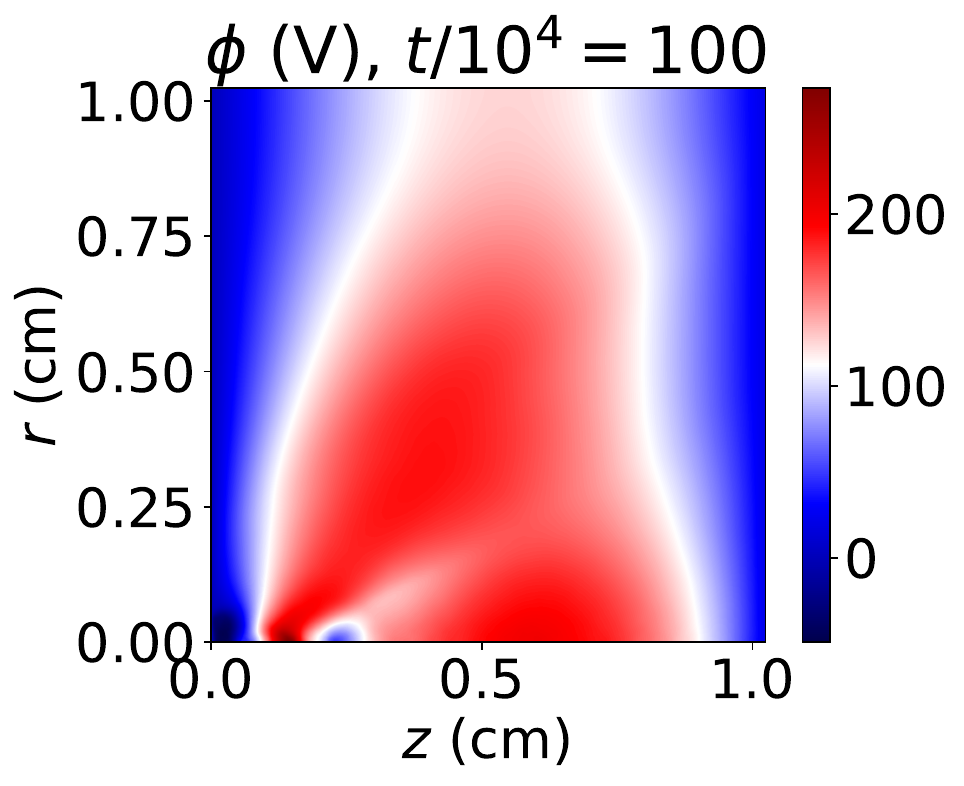}
\includegraphics[height=2.7cm]{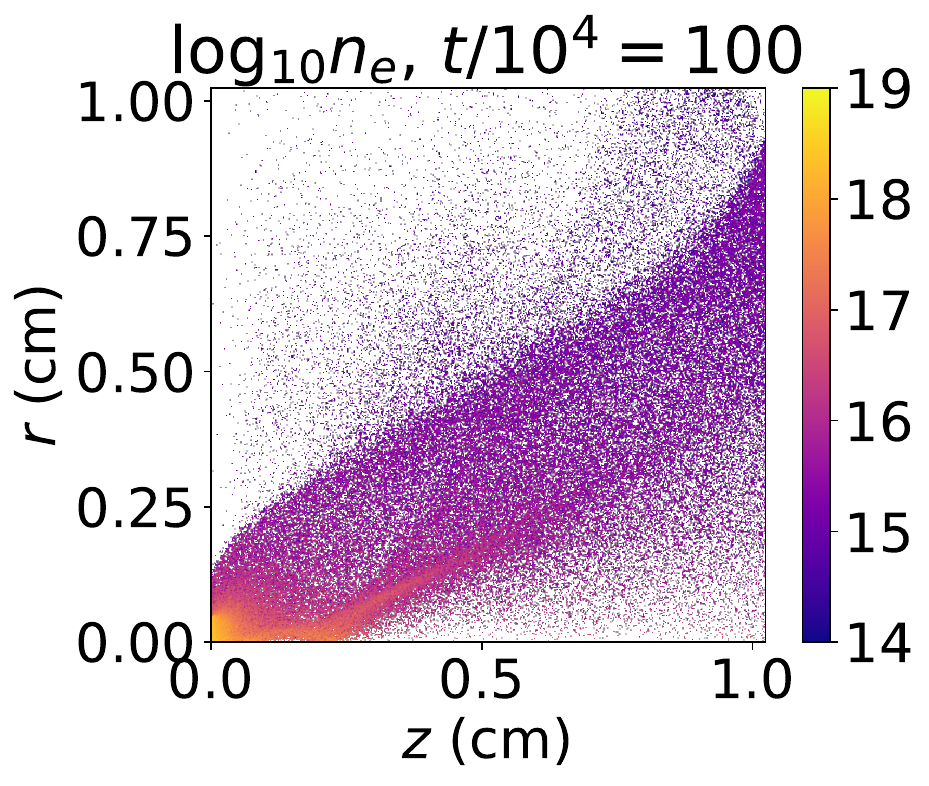}
\includegraphics[height=2.7cm]{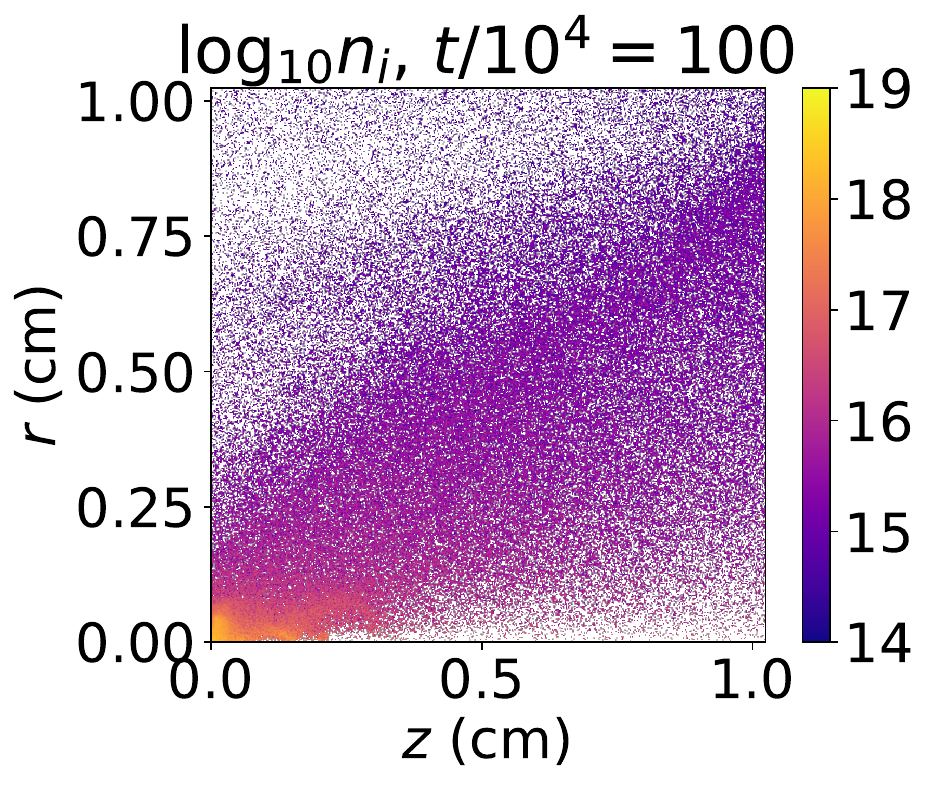}
\includegraphics[height=2.7cm]{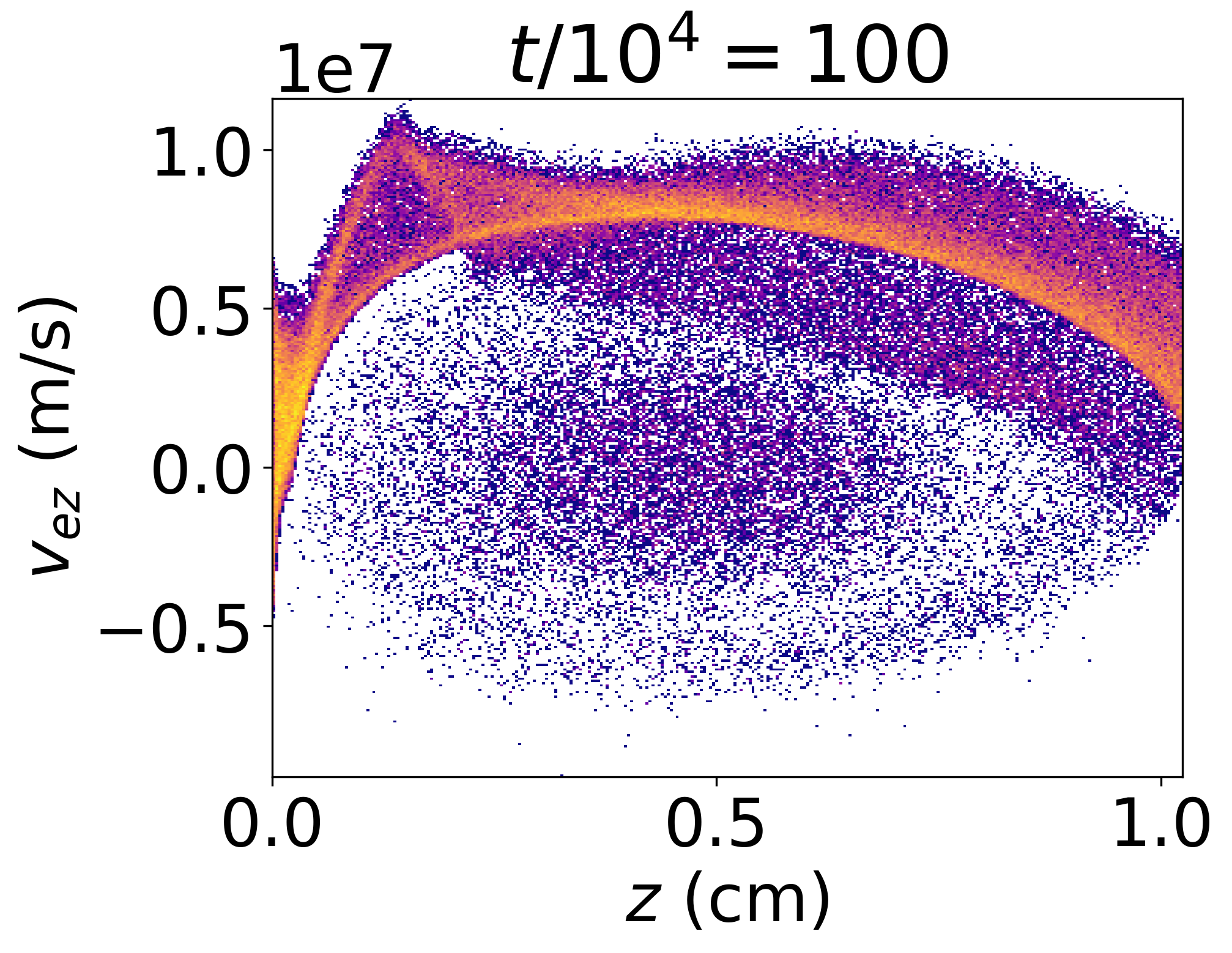}
\includegraphics[height=2.7cm]{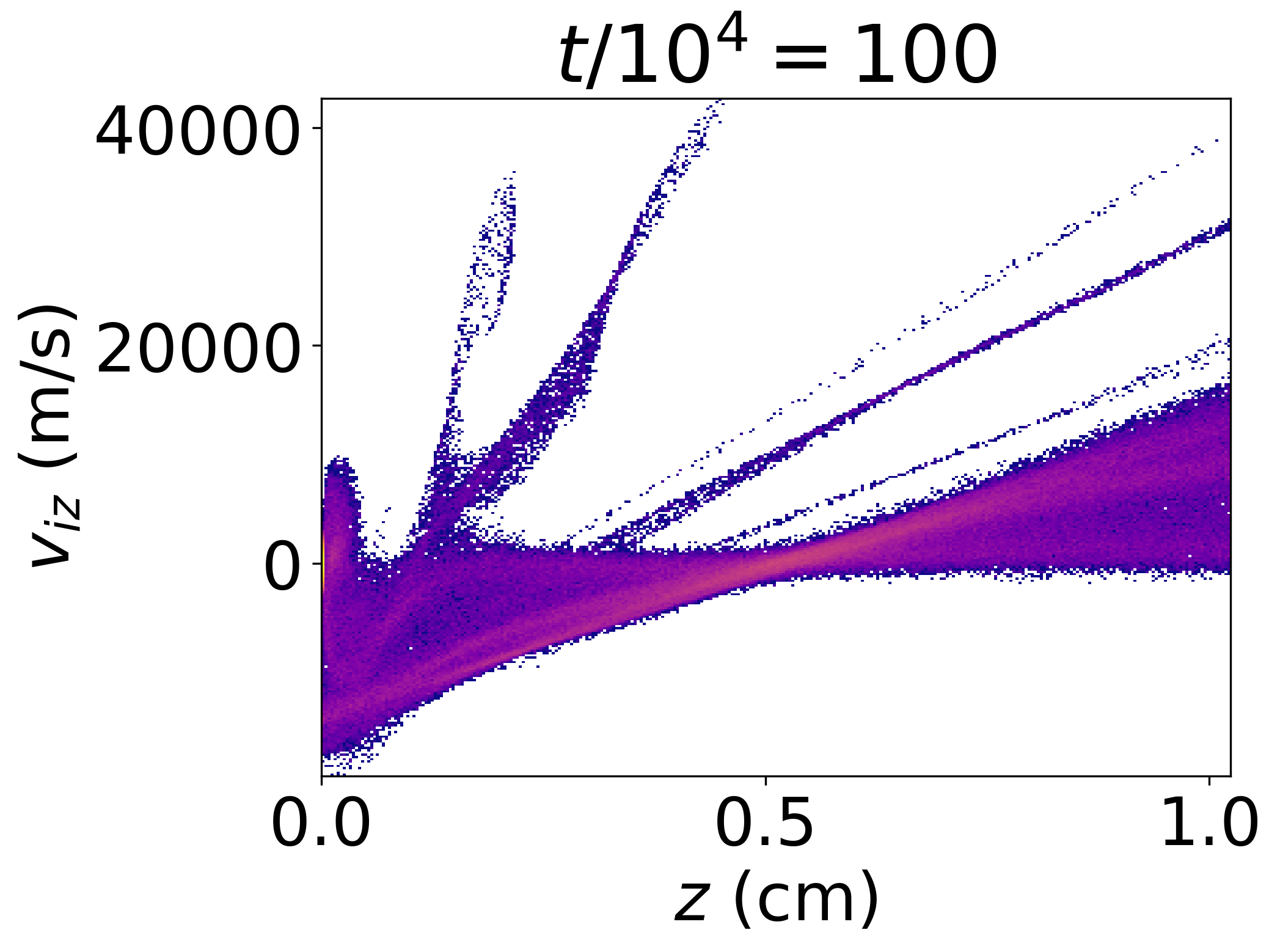}\\
\caption{The time evolution of the charge separation instability
illustrated by the distributions of potential $\phi$, electron density $n_e$ ($m^{-3}$),
ion density $n_i$ ($m^{-3}$), electron and ion $z$-$v_z$ phase space plots,
at different time steps $t$.
\label{fig:process}}
\end{figure*}

\begin{figure*}
\includegraphics[height=2.7cm]{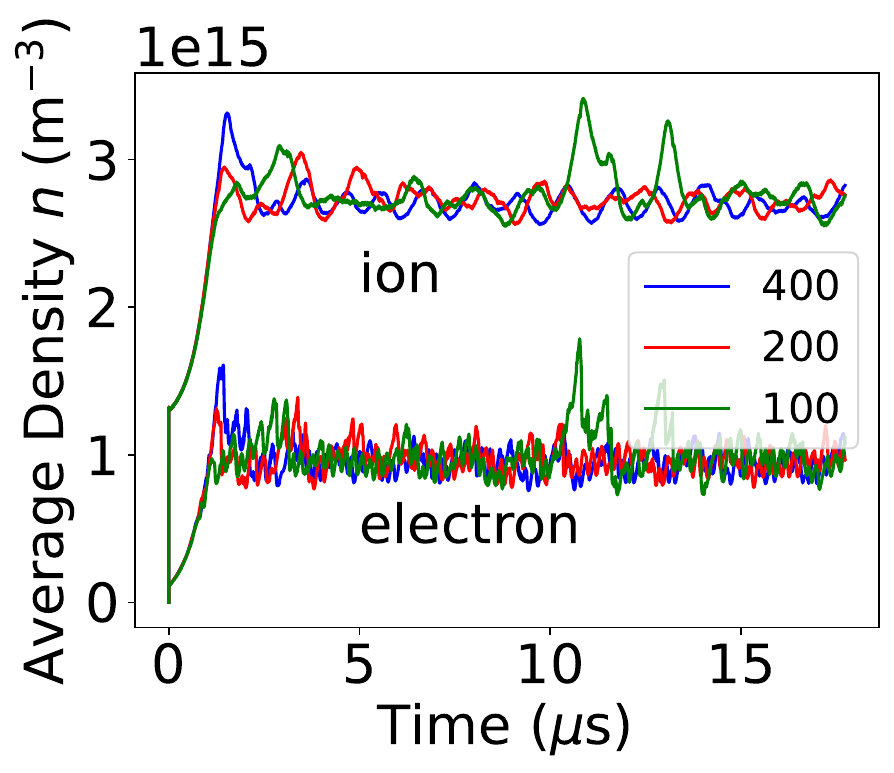}
\includegraphics[height=2.7cm]{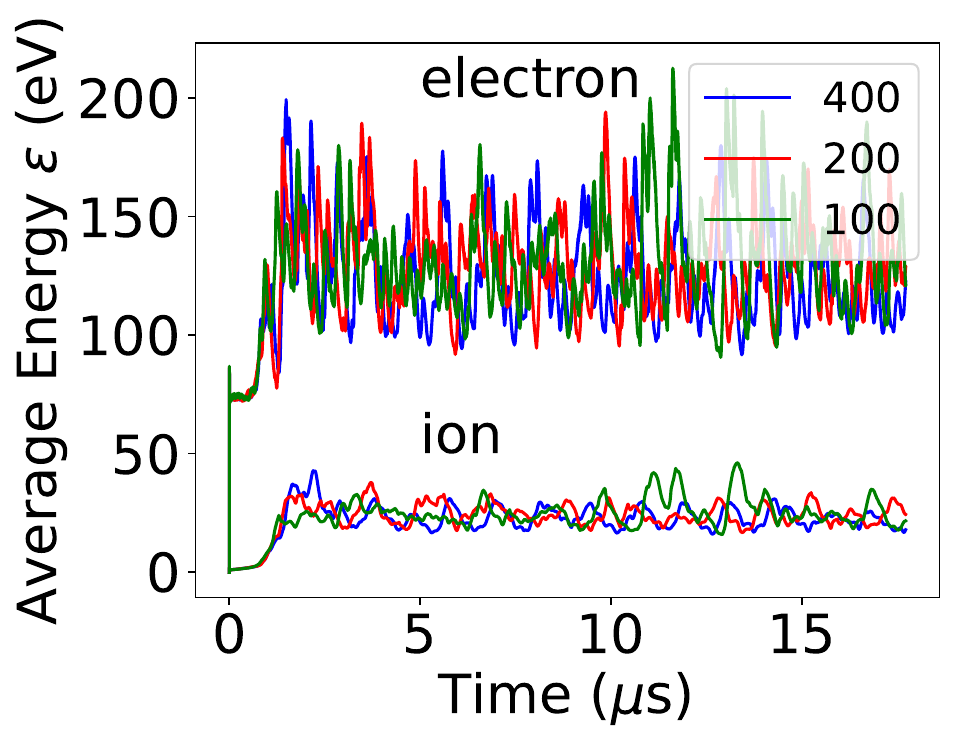}
\includegraphics[height=2.7cm]{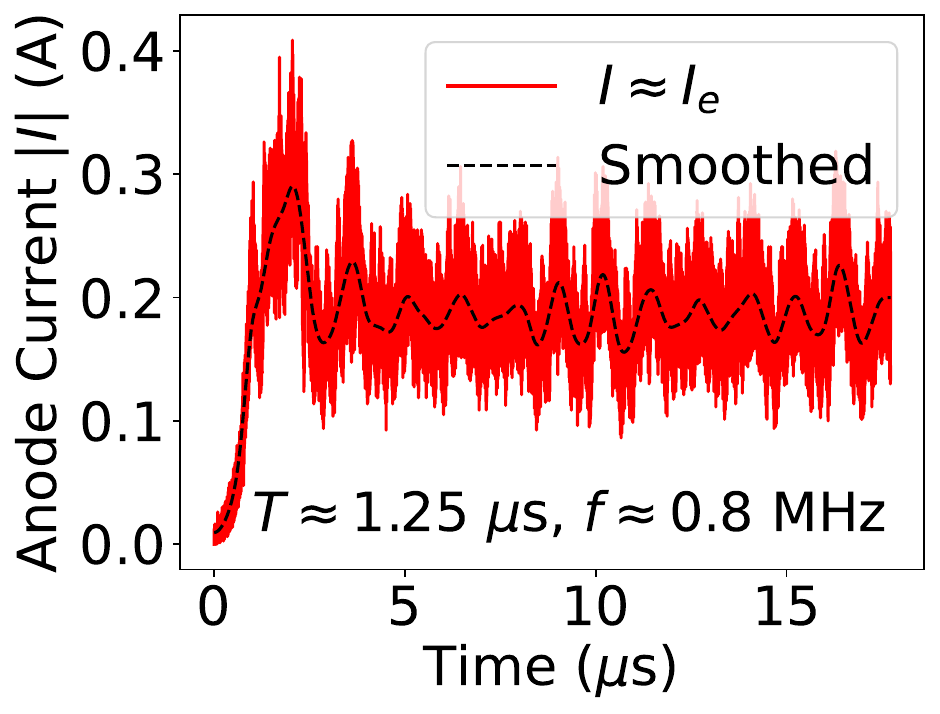}
\includegraphics[height=2.7cm]{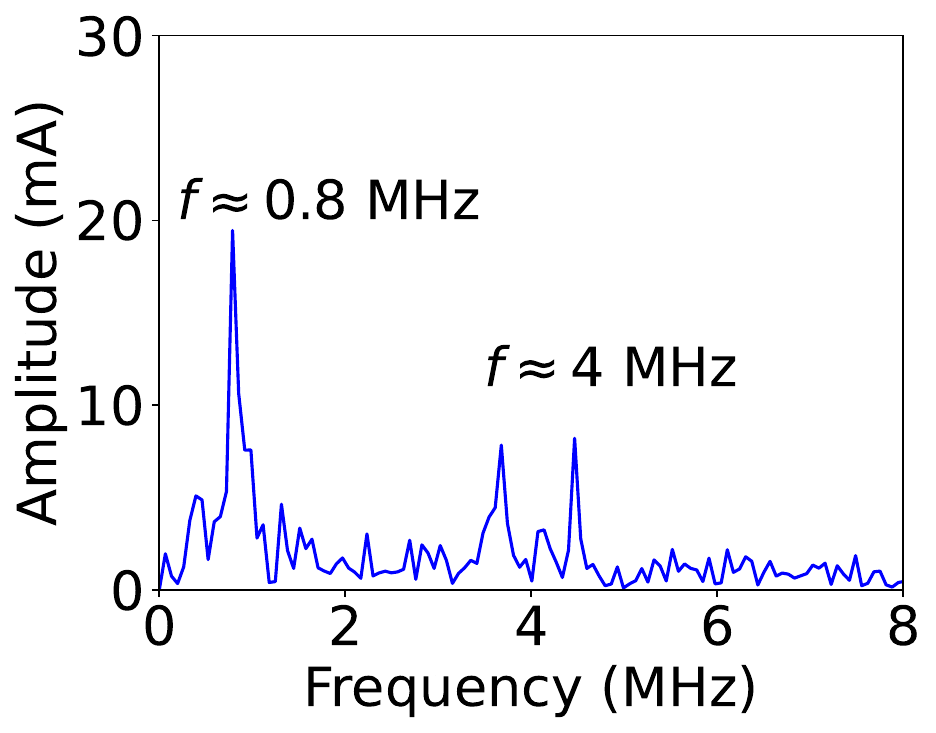}
\includegraphics[height=2.7cm]{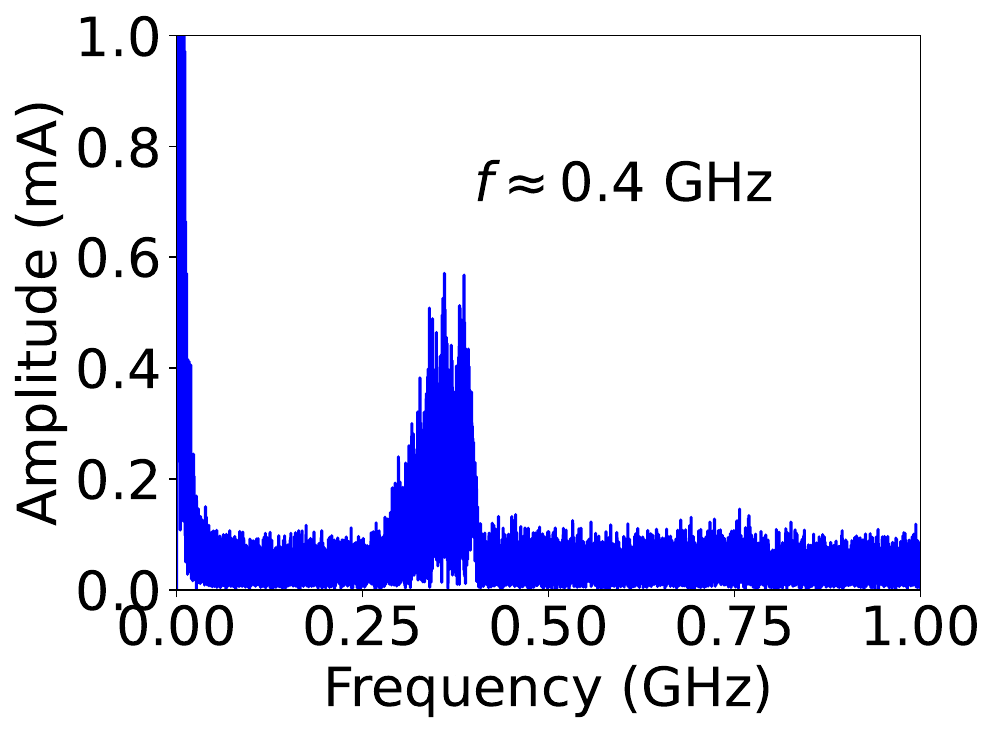}
\caption{
From the left to the right:
the average density and energy over time;
the anode current oscillation over time;
the spectral plot in the MHz scale;
and the spectral plot in the GHz scale.
\label{fig:PE-PNr}}
\end{figure*}

The evolution of the charge separation instability
is captured by the simulation as presented in
Fig.\ref{fig:process}.
(1) At the very beginning,
at time step $t=10^4$ ($\approx 35.5$ ns),
shown in the first row of Fig.\ref{fig:process},
electrons are injected and can travel to the anode,
while expanding axially and radially,
such that the electron density $n_e$ drops quickly.
Because ions are slow
and impeded by the anode potential,
an electron and ion charge separation is formed
and a negative potential ($\sim -100$ V) is established
by those extracted electrons near the cathode exit,
shown in the potential $\phi$ plot.
Looking at the $z$-$v_{iz}$ plot,
there are two groups of ions,
one occupies the whole plume region
with a low density $n_i$,
which are those newly ionized ions;
ions in the other group are extracted
and gain high velocities
up to $v_{iz}\approx12500$ m/s near the cathode exit,
which are pulled by electrons
and try to retrieve the neutrality.
(2) As shown in the second row
of Fig.\ref{fig:process} at
$t=10^5$ ($\approx 0.355$ $\mu$s),
as more ions are extracted by the electrons,
a relatively high potential region ($\sim -30$ V) is formed
at $z \approx 0.1$ cm near the axis,
which is located in between of two
low potential regions,
a small one on the left and a big one on the right.
Looking at the $z$-$v_{iz}$ plot,
when $z<0.1$ cm,
some high speed ions are decelerated
due to the rise of the local potential,
and when $z>0.1$ cm,
some ions are accelerated due to the drop
of the local potential.
The opposite is found in
the $z$-$v_{ez}$ plot for some electrons,
where they are accelerated at $z<0.1$ cm
and decelerated at $z>0.1$ cm.
(3)
Then, there are more extracted ions
to form a higher potential region
($\sim 150$ V),
as shown in the third row
of Fig.\ref{fig:process} at
$t=2.2\times 10^5$ ($\approx 0.781$ $\mu$s).
The previous small low potential region
on the left grows big too,
which is a newly formed group
of electrons,
that help to extract injected ions
and impede injected electrons,
such that some electrons cannot be extracted
and may return back to the cathode,
as shown by the vortex near the
cathode in the $z$-$v_{ez}$ plot.
Looking at the $n_e$ plot,
we can see two high-density branches in the plume,
indicating the discontinuity
of electron extraction.
(4)
The above process reaches a steady-state
after about $2$ $\mu$s,
which can be seen from the average density
and energy plots
in Fig.\ref{fig:PE-PNr},
but the charge separation instability remains,
causing the continuous formation of the
alternating high and low potential regions,
which propagate towards the anode 
along with groups of non-neutral electrons or ions,
as shown by the
last two rows
of Fig.\ref{fig:process}
at time $t=7\times10^5$ ($\approx 2.485$ $\mu$s)
and $t=10^6$ ($\approx 3.55$ $\mu$s).
From the two $z$-$v_{ez}$ plots,
we can see that the center round-shape
electrons are due to ionization,
while those with much higher energy
are mostly formed and driven by the charge separation instability.
From the two $z$-$v_{iz}$ plots,
we can see that after gaining a significant amount
of energy, ions are gradually decelerated
towards the anode as branches one after another.
Those low energy ions in the middle of the plot
are mostly generated by ionization,
and some may move to the anode,
while others may return back to the cathode.

\begin{figure*}
\includegraphics[height=3.0cm]{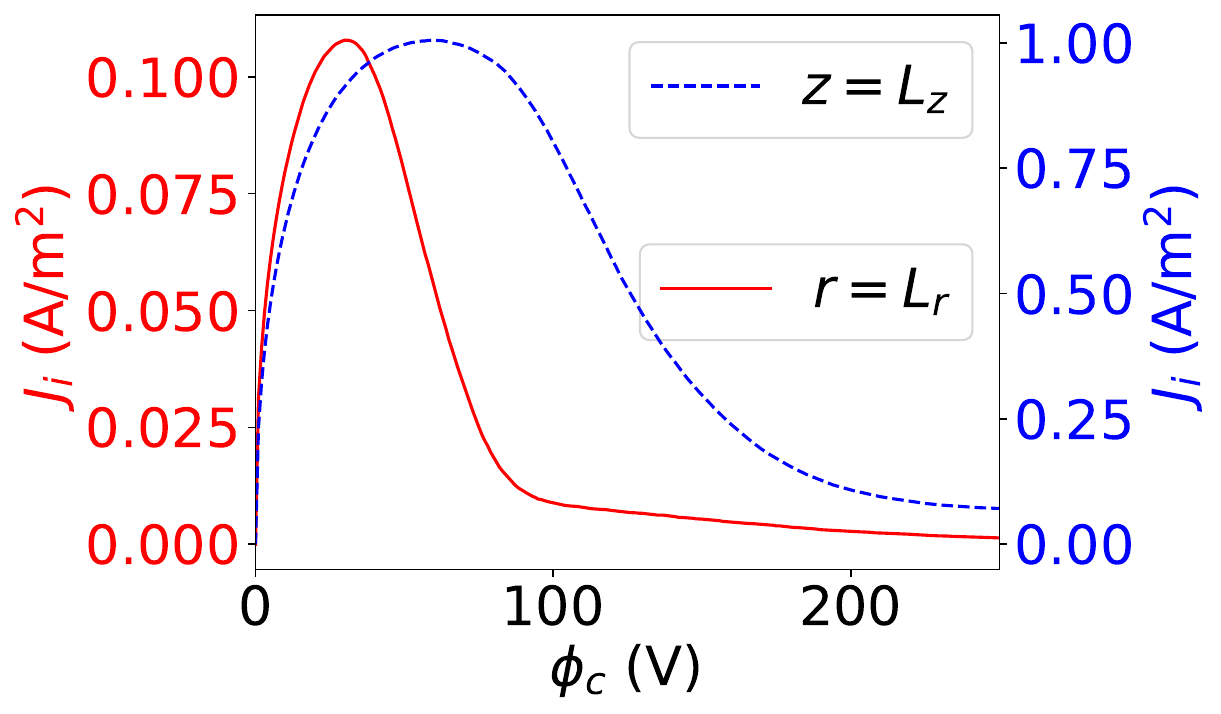}
\includegraphics[height=3.0cm]{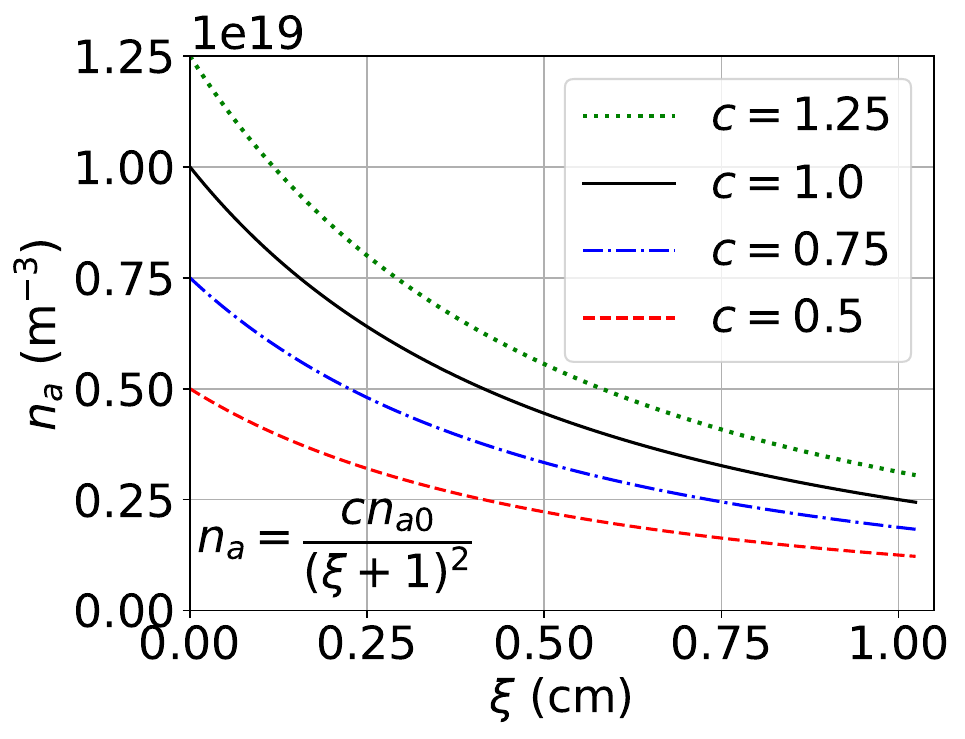}
\includegraphics[height=3.0cm]{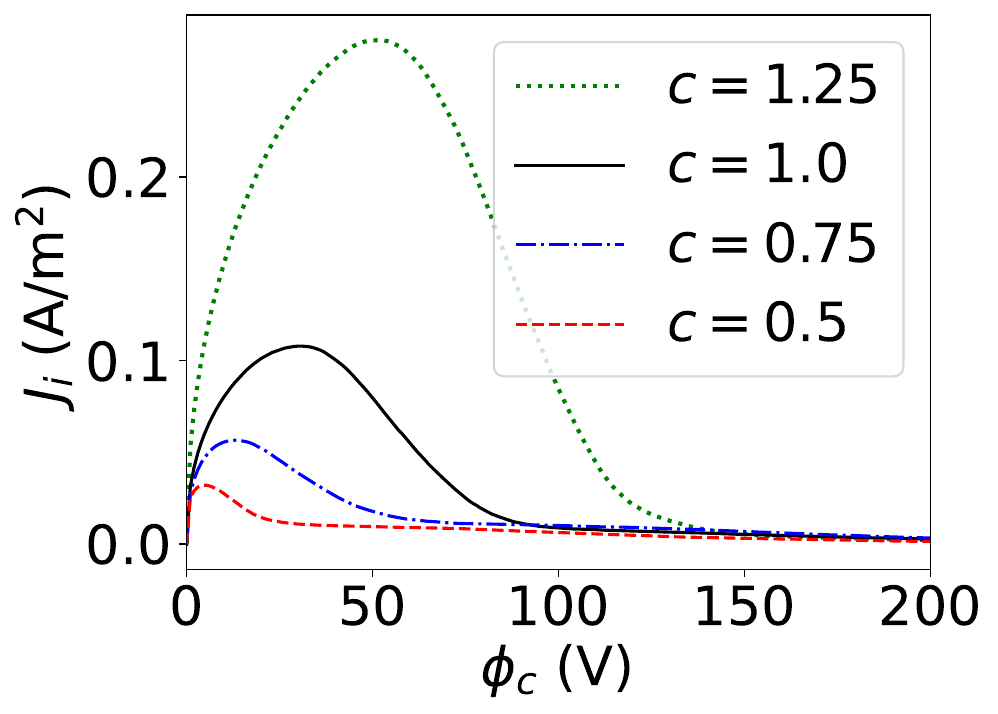}
\includegraphics[height=3.0cm]{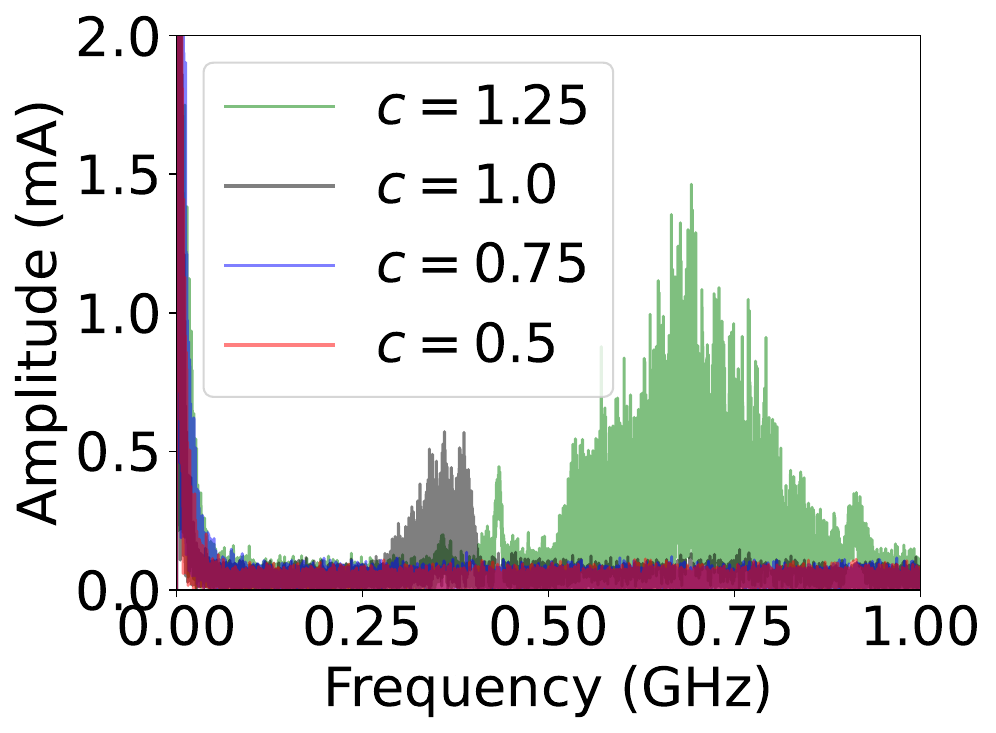}
\caption{
From left to right:
ion current density $J_i$ over collector potential $\phi_c$
on $r=L_r$ and $z=L_z$ two planes;
variation of the neutral atom density distribution $n_a$,
where $\xi^2 = r^2+z^2$;
$J_i$-$\phi_c$ on $r=L_r$ plane
and the GHz spectral plots
of cases varying $n_a$.
\label{fig:fv_na}}
\end{figure*}

As mentioned earlier,
the convergence of the simulation can be seen
in the average density and energy plots
over time, as shown in Fig.\ref{fig:PE-PNr}.
In order to verify that the number of macro-particles
applied is large enough,
two corresponding tests are run with
reduced $N_p=200$ and 100,
and their average density and energy results
are also plotted along with the curves of the base case with $N_p=400$.
As we can see from the figures,
the results match with each other very well.
At the steady state, the average ion density in the plume
is about $2.8 \times 10^{15}$ m$^{-3}$,
and the electron density is about $1 \times 10^{15}$ m$^{-3}$.
The average ion kinetic energy is about 25 eV,
and that for electrons is about 140 eV.
In addition, oscillations are observed in
these average density and energy plots.
To better analyze these oscillations,
and to be more related to diagnostic approaches in experiments,
the number of macro-particles reaching
the anode plane is recorded and
computed to obtain the anode current,
shown in Fig.\ref{fig:PE-PNr},
along with a dashed smoothed curve to
indicate the oscillation period and frequency.
By simply counting the number of oscillation peaks
between a certain time range,
an oscillation with period 1.25 $\mu$s and frequency 0.8 MHz
can be detected.
Further more, FFT is done on the anode current data,
and the plot in the MHz scale shows the 0.8 MHz wave
with an amplitude about 20 mA,
as well as a 4 MHz wave with an amplitude about 10 mA.
These MHz oscillations have been found in experiments too,
such as Fig.12 in \cite{10.1063/5.0188988}
and Fig.15 in \cite{10.1063/1.2784460}.
In the GHz scale,
a 0.4 GHz high frequency wave can be detected with
small amplitude about 0.6 mA,
which was detected too in experiments, such as
Fig.10 in \cite{IEPC-2015-489}.

Then, to diagnose energetic ions,
we mimic the way of RPA measurements commonly used in experiments.
One RPA is considered to be placed on the $r = L_r$ plane,
and those ions within the cylindrical shell with radius
$L_r$, thickness $2\Delta r$, and length $L_z$
are taken into account.
Similarly, another RPA is placed on the $z=L_z$ plane,
and those ions within the cylinder with radius $L_r$
and length $2\Delta z$ are taken into account.
Then, the ``collector'' potential $\phi_c$
is scanned from 0 to 250 V,
and those ions with higher energies
(kinetic energy normal to the RPA surface)
are counted accumulatively,
multiplied by the ion radial/axial velocity, the density,
and the unit charge,
thus the current density $J_i$ can be obtained
as a function of the collector potential $\phi_c$.
The plot is shown in Fig.\ref{fig:fv_na},
from which we can see the $J_i$ collected
has a very similar profile to
many experimental measurements,
such as Fig.2 and Fig.5 in \cite{doi:10.2514/3.23526},
Fig.15 and Fig.16 in \cite{IEPC-2013-076}.
Energetic ions ($> 50$ eV)
are easily detected as those in experiments,
indicating the easiness of energetic ion generation
due to the charge separation instability.

At last, we show that the magnitude of the
neutral atom density distribution $n_a$
can greatly affect the generation of energetic ions.
Different $n_a$ distributions are considered by varying the
constant $c$ in the formula
$n_a = c n_{a0} / (\xi+1)^2$,
where $\xi^2 = r^2 + z^2$,
as shown in Fig.\ref{fig:fv_na}.
The simulation results of the radial
current density $J_i$ for different cases are plotted
in Fig.\ref{fig:fv_na} too,
we can see that decreasing $n_a$ leads to
lower ion energies collected,
which makes sense because as $n_a$ gradually decreases to some point,
the discharge can not be sustained
and no more ions can be generated.
However,
the opposite trend has been found in experiments as well that
decreasing $n_a$ can lead to higher ion energies \cite{Wang_2022}.
This contradiction may be explained
that those experiments and the simulations in this work
are in different regimes,
such as high current (a few amperes) in experiments
versus only 0.5 A in the simulation for saving computational cost.
For the MHz oscillations,
although they are affected by $n_a$,
but no obvious trend can be obtained yet.
For the GHz oscillations,
it can be seen in Fig.\ref{fig:fv_na},
as $n_a$ decreases
GHz oscillations gradually disappear.

In summary,
a new mechanism of charge separation instability
in hollow cathode plume
is found via fully kinetic PIC simulations,
which can easily lead to energetic ions and is
agreed with experimental measurements in the literature
qualitatively.
It is believed that this new finding opens up a
promising avenue to explain the gap between
theories and experiments on the generation of anomalously energetic ions
and the long lasting mystery of hollow cathode instabilities.
Numerous follow-up works can be carried out,
such as more massive simulations with higher currents and larger spatial size,
and parametric studies compared to experiments quantitatively.

\begin{acknowledgments}
The authors acknowledge the support from
National Natural Science Foundation of China
(grant number U22B20120 and 5247120164).
This research used the open-source PIC code
WarpX
%\url{https://github.com/ECP-WarpX/WarpX},
we acknowledge all WarpX contributors.
\end{acknowledgments}

\appendix

%\section{Appendixes}

\nocite{*}

\bibliography{apssamp}% Produces the bibliography via BibTeX.

\end{document}